\documentclass{ws-ijmpd}
\usepackage[super,compress]{cite}
\begin{document}

\markboth{X.-M. Deng}
{The second post-Newtonian Light Propagation and Its Astrometric Measurement in the Solar System}

%
\catchline{}{}{}{}{}
%

\title{THE SECOND POST-NEWTONIAN LIGHT PROPAGATION AND ITS ASTROMETRIC MEASUREMENT IN THE SOLAR SYSTEM }

\author{Xue-Mei Deng}

\address{Purple Mountain Observatory, Chinese Academy of Sciences\\
Nanjing, 210008,
China\\
xmd@pmo.ac.cn}

\maketitle

\begin{history}
\received{Day Month Year}
\revised{Day Month Year}
\end{history}

\begin{abstract}
The relativistic theories of light propagation are generalized by introducing two new parameters $\varsigma$ and $\eta$ in the second post-Newtonian (2PN) order, in addition to the parameterized post-Newtonian parameters $\gamma$ and $\beta$.
This new 2PN parameterized (2PPN) formalism includes the non-stationary gravitational fields and the influences of all kinds of relativistic effects. The multipolar components of gravitating bodies are taken into account as well at the first post-Newtonian order. The equations of motion and their solutions of this 2PPN light propagation problem are obtained. Started from the definition of a measurable quantity, a gauge-invariant angle between the directions of two incoming photons for a differential measurement in astrometric observation is discussed and its formula is derived. For a precision level of a few microacrsecond ($\mu$as) for space astrometry missions in the near future, we further consider a model of angular measurement, LATOR-like missions. In this case, all terms with aimed at the accuracy of $\sim1\mu$as are estimated.
\end{abstract}


\ccode{PACS numbers:04.50.-h, 04.25.Nx, 04.80.Cc}

\allowdisplaybreaks

\section{Introduction}	

With the development of observational techniques and the
improvement of measurement methods, it is time for astrometry to unfold
a new era indubitably. Presently, the satellite laser ranging like laser geodynamics satellite
(LAGEOS) has achieved accuracies of about one millimeter \cite{cp04}, the precision of lunar laser ranging (LLR) has approached one
millimeter \cite{mu08}, and the very long baseline interferometry (VLBI) has
attained to the precision of 0.1 mas or even better \cite{fk09}. Beyond the
current stages, astrometric observation will be able to attain the accuracy of a few
microarcsecond ($\mu$as) or higher for some astrometric missions in the future, such
the Gaia \cite{per01}, the Space Interferometry Mission (SIM) \cite{sha98},
the Square Kilometer Array (SKA) \cite{cr04}, the Laser Astrometric
Test Of Relativity (LATOR) \cite{tsn04} and the Beyond
Einstein Advanced Coherent Optical Network (BEACON) \cite{tsg09}.
This tendency requires a practical framework that can satisfy the accuracy of $\mu$as.

In order to realize this purpose, there are three approaches: post-Newtonian (PN) method,
post-Minkowskian (PM) method and Synge's world function (SWF) method.
Several authors have investigated the light propagation in the
general relativity (GR) and alternative theories of gravity by PN method, which assumes gravitational fields are weak and motions are slow. Richter \&
Matzner \cite{rm82} studied 2PN light deflection for one body in parameterized
post-Newtonian (PPN) formalism \cite{wn72,nw72} by introducing
another parameter $\Lambda$ at $c^{-4}$ of $g_{ij}$ ,
the space-space component of metric. Hellings \cite{hel86} discussed
the relativistic effects in astronomical timing measurements in the PPN formalism, which
could be applied to VLBI, spacecraft ranging and pulsar timing in the gravitational field of
a motionless body. Brumberg
\cite{bru91} detailedly researched the 2PN light propagation of the Schwarzschild solution in three
gauges (standard, harmonic and isotropic) through introducing two coordinate parameters. Brumberg \cite{bru91} also studies the angle between two incoming rays in a motionless $N$-body
system in 1PN approximation of GR. Klioner \& Kopeikin \cite{kk92} developed a
practical relativistic model for the 2PN light propagation in the harmonic gauge of GR, in which the 1PN contributions from bodies in uniform motion and the 2PN
contributions from the Sun were considered. Motivated by ESA's Atomic Clock Ensemble in Space (ACES) \cite{sv96}
on International Space Station (ISS), Blanchet et al. \cite{bst01} discussed time
delay and frequency shift for one body in the isotropic gauge at 1PN of the framework
of general relativity. Klioner \cite{kli03} presented a practical
relativistic model for Gaia under 1PN formalism with PPN parameters $\beta$ and $\gamma$
for uniformly moving bodies. Xie \& Huang \cite{xh08} deduced the 2PN approximation
of Einstein-aether (${\AE}$) theory
in the form of both superpotentials and an $N$-point mass. Kopeikin \cite{kop09}
derived an explicit Lorentz-invariant solution of the Einstein and null geodesic equations for
data processing of the time delay and ranging experiments for a moving body in PPN formalism by introducing
a parameter $\delta$ in spatial isotropic term of $c^{-4}$ for $g_{ij}$.
Minazzoli \& Chauvineau \cite{mc09} extended the IAU2000 resolutions to
include all the $c^{-4}$ terms for the requirements from some
space missions in GR. And then they \cite{mc11} investigated scalar-tensor propagation of light in the inner Solar System including relevant $c^{-4}$ contributions for ranging and time transfer. Klioner \& Zschocke \cite{kz10} formulated the light
propagation in 2PN framework of a stationary gravitational field for the Schwarzschild
metric (one body) in harmonic gauge by introducing one parameter $\epsilon_{\mathrm{kz}}$ in spatial isotropic
and anisotropic terms of $c^{-4}$ for $g_{ij}$. Deng \& Xie \cite{dx12} investigated 2PN light propagation
model at the $c^{-4}$ level in the framework of the scalar-tensor (ST)
theory for a gravitational $N$-point mass system. And they \cite{dx12} found the parametrization with single parameter in spatial isotropic
and anisotropic terms of $c^{-4}$ for $g_{ij}$ is not valid for the ST theory.

Within PM method, it only assumes the condition of weak gravitational fields, but not necessarily slow motion, and it expanded quantities in powers of the
gravitational constant $G$. Kopeikin \& Sch\"{a}fer \cite{ks99} derived the light propagation at the first PM (1PM) linearized Einstein equations in the gravitational field of an arbitrary-moving
$N$-body system. Kopeikin \& Mashhoon \cite{km02} researched the effects of bodies' spins
on the light propagation in the gravitational field of an arbitrary-moving $N$-body system
in the harmonic gauge to 1PM order under GR. A model of the celestial sphere and of the observables for next-generation astrometric missions was
constructed by numerical simulations with the 1PM method in GR for $N$-body of the
Solar System in two cases: static \cite{fcv04} and dynamical \cite{fvc06}. Kopeikin \& Makarov \cite{km07}
took into account light deflection by a giant planet with
multipoles in 1 $\mu$as astronomical measurements using the 1PM method for two-body (the Sun
and Jupiter or Saturn).

In the method of SWF for the light propagation, the integration of geodesic equations can be avoided \cite{lt02,llt04,lt08,tl08,hbl14,hbl14b,bm14}.
Linet \& Teyssandier \cite{lt02} derived time transfer and frequency shift to the order $c^{-4}$ in the field of an axisymmetric rotating body with this method.
Le Poncin-Lafitte, Linet \& Teyssandier \cite{llt04} worked out a recursive procedure for expanding SWF method into a perturbative series of ascending powers of $G$. Le Poncin-Lafitte \& Teyssandier \cite{lt08} used SWF to investigate influence of mass multipolar moments on the deflection of a light ray by an isolated axisymmetric body. Teyssandier \& Le Poncin-Lafitte \cite{tl08} derived the PM expansion of time transfer functions with SWF method. Recently, with the some method (SWF), Hees, Bertone \& Le Poncin-Lafitte \cite{hbl14} present a procedure to compute the relativistic coordinate time delay, Doppler and astrometric observables up to the second PM (2PM) order, and study light propagation in the field of a moving axisymmetric body for the Juno mission \cite{hbl14b}.
Bertone \& Minazzoli et al \cite{bm14} built the time of flight and tangent vectors in a closed form within the SWF formalism giving the case of a time-dependent metric and shown how to use this new approach to obtain a comparison of the time transfer with space astrometric modelings.

\begin{table}[ph]
\tbl{Differences among the metric between
some previous works and our work for light propagation. These include the
order of expansion, whether or not to be parametrized, whether or
not to consider spin, quadrupole moment, $N$-body and bodies' motion.}
{\begin{tabular}{@{}cccccccc@{}} \toprule
Ref.
        & Parametrized & Parameter& Parameter
         &$N$-body & Bodies' & Quadrupole
        &Spin\\
        &  & in $c^{-2}$ & in $c^{-4}$
         & & motions & moment
        &\\ \colrule
\cite{rm82} & Yes &  $\gamma$ & $\beta,\Lambda$&No&No &Yes&Yes\\
\cite{hel86} & Yes & $\gamma$ & $\beta$ & No & No & No & No\\
\cite{bru91} & No &  $--$ & $--$ & No & No & No & No\\
\cite{bru91} & No &  $--$ & $--$ & Yes & No & No & No\\
\cite{kk92} & No &  $--$ & $--$ & Yes & Yes & Yes & Yes\\
\cite{bst01} & No &  $--$ & $--$ & No & No & No & No\\
\cite{kli03} & Yes &  $\gamma$ & $\beta$ & Yes & Yes & Yes & No\\
\cite{xh08} & No &  $--$ & $c_{14}$ & Yes & No & No & No\\
\cite{kop09} & Yes &  $\gamma$ & $\beta,\delta$ & No & Yes & No & No\\
\cite{mc09} & No &  $--$ & $--$ & No & No & No & No\\
\cite{mc11} & No &  $\gamma$ & $\gamma,\beta$ & No & No & No & No\\
\cite{kz10} & Yes &  $\gamma$ & $\beta,\epsilon_{\mathrm{kz}}$ & No & No & No & No\\
\cite{dx12} & No &  $\gamma$ & $\gamma,\beta$ & Yes & No & No & No\\
\cite{ks99} & No &  $--$ & $--$ & Yes & Yes & No & No\\
\cite{km02} & No &  $--$ & $--$ & Yes & Yes & No & Yes\\
\cite{km07} & No &  $--$ & $--$ & No & Yes & Yes & No\\
\cite{lt02} & Yes &  $\gamma$ & $--$ & No & No & Yes & Yes\\
\cite{llt04} & Yes &  $\gamma$ & $\beta,\delta$ & No & No & No & No\\
\cite{lt08} & Yes &  $\gamma$ & $--$ & No & No & Yes & No\\
\cite{tl08} & Yes &  $\gamma$ & $\beta,\delta$ & No & No & No & No\\
\cite{hbl14} & Yes &  $\gamma$ & $\beta,\epsilon_{\mathrm{hbl}}$ & No & No & No & No\\
\cite{hbl14b} & No &  $--$ & $--$ & No & Yes & Yes & No\\
\cite{bm14} & Yes &  $\gamma$ & $--$ & Yes & Yes & No & No\\
Our work & Yes &  $\gamma$ & $\beta,\varsigma,\eta$ & Yes & Yes & Yes & Yes\\ \botrule
\end{tabular} \label{ta1}}
\end{table}

In this work, we use the PN method. Table \ref{ta1} gives a comparison among these previous works and our present work. Firstly,
it is worth emphasizing that we employ the same integral technique and iterative method in  Refs.~\refcite{bru91,kk92,kli03,kz10}. Aimed at the level of $\mu$as, our work has the following aspects:
\begin{itemlist}
 \item $N$ gravitating bodies in the Solar System, including the effects of their quadrupole moments and spins;
 \item motions of the $N$-body.  Kopeikin \cite{kop09} pointed out and
discussed that these motions would contaminate the observed
numerical value of PPN parameters. In our model, we separate them
into two parts: the 1PN contributions from the rectilinear and
uniform $N$-body (in a short time span) and the 2PN contributions from
the motionless Sun, which is sufficient for near future missions;
 \item two new 2PN parameters $\varsigma$ and $\eta$
in addition to the two PPN parameters $\beta$ and $\gamma$. $\varsigma$ and $\eta$ respectively parametrize the spatial isotropic and anisotropic terms in $g_{ij}$
at $\mathcal{O}(c^{-4})$. In
some cases, $\varsigma$ and $\eta$ are functions of $\beta$ and
$\gamma$ \cite{dx12}, but, in other cases, they can
 be independent
of the PPN parameters \cite{xh08}.
\end{itemlist}

In what follows, our conventions and notations generally follow those of Ref.~\refcite{mtw73}.
The plot of this paper is as follows. With the level of accuracy of $\sim\mu$as, we develop a 2PN parametrized (2PPN) framework for light
propagation in section 2. Subsequently, in section
3, a gauge-invariant angle between incoming light-rays is derived and some special cases are
shown. Then, we discuss a LATOR-like model of angle measurement: an observer measures the angle between two incoming rays emitted separately from two spacecrafts which are all at the Earth's orbit circling the Sun. With the level of accuracy of $\sim1\mu$as, we estimate the magnitudes of contributions from various physical sources
in this case. Finally, conclusion and perspectives are outlined in section 4.

\section{2PPN framework for light propagation}

\subsection{2PPN metric for light}

In our investigation, we neglect the effects of our galaxy and external galaxies on the Solar System. This means the Solar System is an isolated system. Observations are modeled in the Solar System barycentric reference system (SSBRS), which can be mathematically described by a metric tensor.

For practical reasons, we only care about the effects more than $1\mu$as in this work. It means the metric for light can be necessarily (but not over) simplified. The 2PPN metric for SSBRS  reads as
\begin{eqnarray}
\label{metric00}
g_{00}&=&-1+\epsilon^{2}2\sum_{A}\bigg(\frac{Gm_{A}}{r_{A}}+\frac{3}{2}\frac{G}{r^{5}_{A}}J^{<ik>}_{A}r^{i}_{A}r^{k}_{A}\bigg)
-\epsilon^{4}2\beta\frac{G^{2}m^{2}_{\odot}}{r^{2}_{\odot}}
+\mathcal{O}(5),\\
\label{metric0i}
g_{0i}&=&-\epsilon^{3}2(1+\gamma)\sum_{A}\frac{Gm_{A}}{r_{A}}v^{i}_{A}-\epsilon^{3}2(1+\gamma)\sum_{A}\frac{G}{r^{3}_{A}}\varepsilon^{i}_{jk}S^{j}_{A}r^{k}_{A}
+\mathcal{O}(5),\\
\label{metricij}
g_{ij}&=&\delta_{ij}+\epsilon^{2}\delta_{ij}2\gamma\sum_{A}\bigg(\frac{Gm_{A}}{r_{A}}+\frac{3}{2}\frac{G}{r^{5}_{A}}J^{<kl>}_{A}r^{k}_{A}r^{l}_{A}\bigg)\nonumber\\
&&+\epsilon^{4}\bigg\{\delta_{ij}\varsigma\frac{G^{2}m^{2}_{\odot}}{r^{2}_{\odot}}+\eta\frac{G^{2}m^{2}_{\odot}}{r^{4}_{\odot}}r^{i}_{\odot}r^{j}_{\odot}\bigg\}
+\mathcal{O}(5),
\end{eqnarray}
where $\epsilon=1/c$ and $\mathcal{O}(n)$ means of order $\epsilon^{n}$. $m_{A}$ and $S^{i}_{A}$ are the mass and intrinsic angular momentum
(spin) of the body $A$; subscript ``$\odot$" denotes the evaluation
of the Sun; ``$v_{A}^{i}$" is the coordinate velocity of the body
$A$; $r^{i}_{A}=x^{i}-x^{i}_{A}(t)$ and the trajectory of the body
$A$ is represented by $x^{i}_{A}(t)$ and
$r_{A}=(r^{i}_{A}r^{i}_{A})^{1/2}$. And
$J^{<ik>}_{A}=J^{ik}_{A}-\frac{1}{3}\delta^{ik}J^{ss}_{A}$ is the
symmetric trace-free (STF) quadrupole moment of body $A$. $\varsigma$ and $\eta$ are two new 2PPN parameters.

\begin{table}[ph]
\tbl{The order of
$\frac{Gm_{A}}{c^{2}r_{AB}}$ between bodies in the Solar System.
This table shows that the bodies at columns affect the bodies at
rows when a light ray arrives their limbs.}
{\begin{tabular}{@{}cccccccccc@{}} \toprule
$\frac{Gm_{A}}{c^{2}r_{AB}}$ &Sun
        & Mercury & Venus& Earth
         &Mars & Jupiter & Saturn
        &Uranus&Neptune\\ \colrule
Sun&$2.1\times10^{-6}$&$2.6\times10^{-8}$&$1.4\times10^{-8}$&$9.9\times10^{-9}$
&$6.5\times10^{-9}$&$1.9\times10^{-9}$&$1.0\times10^{-9}$&$5.1\times10^{-10}$&$3.3\times10^{-10}$\\
Mercury&$4.2\times10^{-15}$&$1.0\times10^{-10}$&$4.9\times10^{-15}$&$2.7\times10^{-15}$
&$1.4\times10^{-15}$&$3.4\times10^{-16}$&$1.8\times10^{-16}$&$8.7\times10^{-17}$&$5.5\times10^{-17}$\\
Venus&$3.3\times10^{-14}$&$7.2\times10^{-14}$&$6.0\times10^{-10}$&$8.7\times10^{-14}$
&$3.0\times10^{-14}$&$5.4\times10^{-15}$&$2.7\times10^{-15}$&$1.3\times10^{-15}$&$8.2\times10^{-16}$\\
Earth&$3.0\times10^{-14}$&$4.8\times10^{-14}$&$1.1\times10^{-13}$&$7.0\times10^{-10}$
&$5.7\times10^{-14}$&$7.1\times10^{-15}$&$3.5\times10^{-15}$&$1.6\times10^{-15}$&$1.0\times10^{-15}$\\
Mars&$2.1\times10^{-15}$&$2.8\times10^{-15}$&$4.0\times10^{-15}$&$6.1\times10^{-15}$
&$1.4\times10^{-10}$&$8.7\times10^{-15}$&$4.0\times10^{-16}$&$1.8\times10^{-16}$&$1.1\times10^{-16}$\\
Jupiter&$1.8\times10^{-12}$&$2.0\times10^{-12}$&$2.1\times10^{-12}$&$2.2\times10^{-12}$
&$2.6\times10^{-12}$&$2.0\times10^{-8}$&$2.2\times10^{-12}$&$6.7\times10^{-13}$&$3.8\times10^{-13}$\\
Saturn&$3.0\times10^{-13}$&$3.1\times10^{-13}$&$3.2\times10^{-13}$&$3.3\times10^{-13}$
&$3.5\times10^{-13}$&$6.5\times10^{-13}$&$7.0\times10^{-9}$&$2.9\times10^{-13}$&$1.4\times10^{-13}$\\
Uranus&$2.3\times10^{-14}$&$2.3\times10^{-14}$&$2.3\times10^{-14}$&$2.4\times10^{-14}$
&$2.4\times10^{-14}$&$3.1\times10^{-14}$&$4.5\times10^{-14}$&$2.5\times10^{-9}$&$4.0\times10^{-14}$\\
Neptune&$1.7\times10^{-14}$&$1.7\times10^{-14}$&$1.7\times10^{-14}$&$1.8\times10^{-14}$
&$1.8\times10^{-14}$&$2.0\times10^{-14}$&$2.5\times10^{-14}$&$4.7\times10^{-14}$&$1.4\times10^{-9}$\\ \botrule
\end{tabular} \label{ta2}}
\end{table}

\begin{table}[ph]
\tbl{The relativistic effect of the oblateness of a body on light at its surface,
$\frac{Gm_{A}}{c^{2}a_{eA}}J_{2A}$.
Where $m_{A}$, $a_{eA}$ and $J_{2A}$ are respectively the mass, the equatorial radius and the dynamical
form factor for body $A$.}
{\begin{tabular}{@{}ccccccccc@{}} \toprule
Sun
        & Mercury & Venus& Earth
         &Mars & Jupiter & Saturn
        &Uranus&Neptune\\ \colrule
$2.1\times10^{-13}$  & $6.0\times10^{-15}$  & $2.4\times10^{-15}$   & $7.5\times10^{-13}$  & $2.8\times10^{-13}$ &
$2.9\times10^{-10}$ & $1.4\times10^{-11}$ & $8.4\times10^{-12}$ & $5.2\times10^{-12}$\\ \botrule
\end{tabular} \label{ta3}}
\end{table}

At the 1PN order, we keep all of the contributions from the mass monopoles of the Sun and the planets in the $\epsilon^2$ terms of $g_{00}$ and $g_{ij}$. Table \ref{ta2} shows their effects of the major bodies in the Solar System on the light ray when the light ray grazes the limb of one body. The biggest one comes from the Sun, which is about $10^{-6}$. We also include the quadrupole moments at this order to take account their leading contributions. Table \ref{ta3} shows the effect of non-spherical part ($J_{2}$) of a body on light when a light ray grazes its limb. Although the fourth ($J_{4}$) and the sixth ($J_{6}$) zonal spherical harmonics for the giant planets, such as Jupiter and Saturn, can cause deflections larger than 1 $\mu$as for light deflection when the light ray arrived their limbs \cite{sof03}, we leave them untouched in this work. In fact, their effects can be easily taken into account from the studies for Gaia mission \cite{kli03,kz10}.

\begin{table}[ph]
\tbl{The order of
$\frac{Gm_{A}}{c^{3}r_{AB}}v_{A}$ between bodies in the Solar
System. This table shows that the bodies at columns affect the
bodies at rows when a light ray arrives their limbs.}
{\begin{tabular}{@{}cccccccccc@{}} \toprule
$\frac{Gm_{A}}{c^{3}r_{AB}}v_{A}$ &Sun
        & Mercury & Venus& Earth
         &Mars & Jupiter & Saturn
        &Uranus&Neptune\\ \colrule
Sun&$1.1\times10^{-13}$&$1.3\times10^{-15}$&$6.8\times10^{-16}$&$4.9\times10^{-16}$
&$3.2\times10^{-16}$&$9.5\times10^{-17}$&$5.2\times10^{-17}$&$2.6\times10^{-17}$&$1.6\times10^{-17}$\\
Mercury&$6.8\times10^{-19}$&$1.6\times10^{-14}$&$7.8\times10^{-19}$&$4.3\times10^{-19}$
&$2.3\times10^{-19}$&$5.4\times10^{-20}$&$2.9\times10^{-20}$&$1.4\times10^{-20}$&$8.8\times10^{-21}$\\
Venus&$3.9\times10^{-18}$&$8.4\times10^{-18}$&$7.0\times10^{-14}$&$1.0\times10^{-17}$
&$3.5\times10^{-18}$&$6.3\times10^{-19}$&$3.2\times10^{-19}$&$1.5\times10^{-19}$&$9.6\times10^{-20}$\\
Earth&$2.9\times10^{-18}$&$4.8\times10^{-18}$&$1.1\times10^{-17}$&$6.9\times10^{-14}$
&$5.6\times10^{-18}$&$7.0\times10^{-19}$&$3.5\times10^{-19}$&$1.6\times10^{-19}$&$1.0\times10^{-19}$\\
Mars&$1.7\times10^{-19}$&$2.3\times10^{-19}$&$3.2\times10^{-19}$&$4.9\times10^{-19}$
&$1.1\times10^{-14}$&$7.0\times10^{-20}$&$3.2\times10^{-20}$&$1.5\times10^{-20}$&$9.0\times10^{-21}$\\
Jupiter&$7.9\times10^{-17}$&$8.5\times10^{-17}$&$9.2\times10^{-17}$&$9.8\times10^{-17}$
&$1.1\times10^{-16}$&$8.6\times10^{-13}$&$9.5\times10^{-17}$&$2.9\times10^{-17}$&$1.7\times10^{-17}$\\
Saturn&$9.5\times10^{-18}$&$9.9\times10^{-18}$&$1.0\times10^{-17}$&$1.1\times10^{-17}$
&$1.1\times10^{-17}$&$2.1\times10^{-17}$&$2.3\times10^{-13}$&$9.4\times10^{-18}$&$4.4\times10^{-18}$\\
Uranus&$5.1\times10^{-19}$&$5.2\times10^{-19}$&$5.3\times10^{-19}$&$5.4\times10^{-19}$
&$5.5\times10^{-19}$&$7.0\times10^{-19}$&$1.0\times10^{-18}$&$5.7\times10^{-14}$&$9.0\times10^{-19}$\\
Neptune&$3.1\times10^{-19}$&$3.1\times10^{-19}$&$3.1\times10^{-19}$&$3.2\times10^{-19}$
&$3.2\times10^{-19}$&$3.7\times10^{-19}$&$4.5\times10^{-19}$&$8.5\times10^{-19}$&$2.8\times10^{-14}$\\ \botrule
\end{tabular} \label{ta4}}
\end{table}

In $g_{0i}$ [see Eq. (\ref{metric0i})], we include all of the contributions from the mass monopoles and their velocities of the Sun and the planets in the $\epsilon^3$ terms of $g_{0i}$. Table \ref{ta3} gives the estimation of $g_{0i}$ related to a body's orbital motion $v_{A}^{i}$.  In order to investigate the influence of the gravitomagnetic fields on light propagation, not only a body's orbital motion but also its spin are taken into account in the metric of $g_{0i}$ in our model.

For the 2PN terms in $g_{00}$  [see Eq. (\ref{metric00})] and $g_{ij}$ [see Eq. (\ref{metricij})], we only consider the contributions of the monopole of the Sun and neglect the 2PN contributions due to the planets as well as the nonlinear combinations between the planets and the Sun, and the terms $\mathcal{O}(\epsilon^{4}J_{2A})$ whose contributions are less than $1\mu$as. In the $\epsilon^4$ terms of $g_{00}$ and $g_{ij}$, we also omit the terms explicitly depending on $v^{2}_{\odot}$, such as $\frac{Gm_{\odot}}{c^{4}r_{\odot}}v^{2}_{\odot}$, because their contributions to the deflection are at the level of femto-arcsecond.

It is very important to point out that the $\epsilon^2$ terms in $g_{00}$ and $g_{ij}$ and the $\epsilon^3$ terms in $g_{0i}$ are all time-dependent, even including the terms associated with the Sun due to its barycentric motion $x^{i}_{\odot}(t)$ and $v^{i}_{\odot}(t)$. Kopeikin \cite{kop09} demonstrated that this motion of the Sun affects the measured values of the PPN parameters, and cannot be ignored. Moreover, the barycentric motion of the Sun was crucial in order to test general relativity in the Cassini experiment \cite{kpsv07}. Thus, when we integrate the trajectory of the light-ray (see Sec. 2.3), we will keep the velocities of all the bodies including the Sun.

\begin{table}[ph]
\tbl{$\varsigma$ and $\eta$ at the spatial isotropic and anisotropic
parts of $c^{-4}$ for $g_{ij}$ in harmonic
gauge in different theories. $\beta$ and $\gamma$
 are PPN parameters and $c_{14} = c_{1} + c_{4}$, where $c_{1}$ and $c_{4}$
are constant parameters in {\AE} theory \cite{xh08}.}
{\begin{tabular}{@{}cccc@{}} \toprule
Parameter & GR  & ST  & {\AE}  \\ \colrule
$\varsigma$    & 1            & $2\gamma^{2}-\frac{1}{2}\gamma+2\beta-\frac{5}{2}$         & $1+\frac{1}{2}c_{14}$       \\
$\eta$ & 1    & $\frac{1}{2}(1+\gamma)$        & $1-\frac{1}{2}c_{14}$   \\
Ref. &     & \cite{dx12}        & \cite{xh08}   \\ \botrule
\end{tabular} \label{ta5}}
\end{table}

At the 2PN order of $g_{ij}$, we introduce two new parameters $\varsigma$ and $\eta$, which generalize the previous works. They have different values in different gravitational theories (see Table \ref{ta5}). When $\varsigma=\eta=1$, the  Eq. (\ref{metricij}) reduces to GR. When $\varsigma=2\gamma^{2}-\frac{1}{2}+2\beta-\frac{5}{2}$ and $\eta=\frac{1}{2}(1+\gamma)$, the  Eq. (\ref{metricij}) returns to
ST \cite{dx12}. When $\varsigma=1+\frac{1}{2}c_{14}$  and $\eta=1-\frac{1}{2}c_{14}$, the  Eq. (\ref{metricij}) coincides with {\AE} \cite{xh08}. When $\varsigma=\eta$,
the  Eq. (\ref{metricij}) goes back to the results of Ref.~\refcite{kz10}. When $\varsigma=\frac{3}{2}\delta$ and $\eta=0$,
the  Eq. (\ref{metricij}) coincides with the result of Ref.~\refcite{kop09}.

\subsection{2PPN equations of light}

Generally, for a photon propagating in a spacetime
in which Einstein Equivalence Principle (EEP) is valid, the basic
equations of light \cite{bru91} are
\begin{eqnarray}
g_{\mu\nu}\frac{\mathrm{d}x^{\mu}}{\mathrm{d}\lambda}\frac{\mathrm{d}x^{\nu}}{\mathrm{d}\lambda}&=&0,\\
\frac{\mathrm{d}^{2}x^{\mu}}{\mathrm{d}\lambda^{2}}+\Gamma^{\mu}_{\nu\sigma}\frac{\mathrm{d}x^{\nu}}{\mathrm{d}\lambda}
\frac{\mathrm{d}x^{\sigma}}{\mathrm{d}\lambda}&=&0.
\end{eqnarray}

We replace the affine parameter $\lambda$ with coordinate time $t$,
and the equations become
\begin{eqnarray}
\label{ds}
0&=&g_{\mu\nu}\frac{\mathrm{d}x^{\mu}}{\mathrm{d}t}\frac{\mathrm{d}x^{\nu}}{\mathrm{d}t},\\
\label{light}
\frac{\mathrm{d}^{2}x^{i}}{\mathrm{d}t^{2}}&=&\bigg(\frac{1}{c}\Gamma^{0}_{\nu\sigma}
\frac{\mathrm{d}x^{i}}{\mathrm{d}t}-\Gamma^{i}_{\nu\sigma}\bigg)\frac{\mathrm{d}x^{\nu}}{\mathrm{d}t}\frac{\mathrm{d}x^{\sigma}}{\mathrm{d}t}.
\end{eqnarray}
Assuming $\dot{\mathbf{x}}\equiv cs\bm{\mu}=\mathcal{O}(1)$ and
$\bm{\mu}\cdot\bm{\mu}=1$, we find the expression for $s$ from
Eq. (\ref{ds}). Then, by substituting
$\dot{\mathbf{x}}\cdot\dot{\mathbf{x}}\equiv c^{2}s^{2}$
and the
metric Eqs. (\ref{metric00})-(\ref{metricij}) into Eq.
(\ref{light}), we obtain the 2PPN equations of light
propagation in SSBRS as follows
\begin{equation}
\label{lightequation}
\ddot{x}^{i}=\mathcal{F}^{i}_{1PN}+\mathcal{F}^{i}_{Q}+\mathcal{F}^{i}_{S}+\mathcal{F}^{i}_{2PN},
\end{equation}
where the 1PN monopole component with the orbital motion is
\begin{eqnarray}
\label{PN}
\mathcal{F}^{i}_{1PN}&=&-(1+\gamma)\sum_{A}\frac{Gm_{A}}{r^{3}_{A}}\bigg\{\bigg(1-2\frac{\dot{\mathbf{x}}\cdot\mathbf{v}_{A}}{c^{2}}\bigg)r^{i}_{A}
+2\frac{\dot{\mathbf{x}}\cdot\mathbf{r}_{A}}{c^{2}}v^{i}_{A}+\bigg[\frac{\mathbf{r}_{A}\cdot\mathbf{v}_{A}}{c}\nonumber\\
&&-2\bigg(1-\frac{\dot{\mathbf{x}}\cdot\mathbf{v}_{A}}{c^{2}}\bigg)\frac{\dot{\mathbf{x}}\cdot\mathbf{r}_{A}}{c}\bigg]\frac{\dot{x}^{i}}{c}\bigg\},
\end{eqnarray}
the influence of quadrupole moments of the bodies in the Solar
System is
\begin{eqnarray}
\label{Q}
\mathcal{F}^{i}_{Q}&=&\frac{3(1+\gamma)}{2}\sum_{A}\frac{G}{r^{5}_{A}}J^{<jk>}_{A}r^{k}_{A}
\bigg[-5\frac{r^{i}_{A}r^{j}_{A}}{r^{2}_{A}}+2\delta_{ij}
-2\bigg(2\frac{\dot{x}^{j}}{c}-5\frac{\dot{\mathbf{x}}\cdot\mathbf{r}_{A}}{cr^{2}_{A}}r^{j}_{A}\bigg)\frac{\dot{x}^{i}}{c}\bigg],
\end{eqnarray}
the effect from their spins is
\begin{eqnarray}
\label{R}
\mathcal{F}^{i}_{S}&=&2(1+\gamma)\sum_{A}\frac{G}{r^{3}_{A}}\bigg\{\frac{(\mathbf{S}_{A}\times\dot{\mathbf{x}})^{i}}{c^{2}}
+\frac{3}{2c^{2}r^{2}_{A}}[(\mathbf{S}_{A}\times\mathbf{r}_{A})\cdot\dot{\mathbf{x}}]r^{i}_{A}
-\frac{3}{2c^{4}r^{2}_{A}}[(\mathbf{S}_{A}\times\mathbf{r}_{A})\cdot\dot{\mathbf{x}}](\dot{\mathbf{x}}\cdot\mathbf{r}_{A})\dot{x}^{i}\nonumber\\
&&-\frac{3}{2c^{2}r^{2}_{A}}(\dot{\mathbf{x}}\cdot\mathbf{r}_{A})(\mathbf{S}_{A}\times\mathbf{r}_{A})^{i}\bigg\},
\end{eqnarray}
and the 2PN monopole component of the Sun is
\begin{eqnarray}
\label{PPN}
\mathcal{F}^{i}_{2PN}&=&2\frac{G^{2}m^{2}_{\odot}}{c^{2}r^{4}_{\odot}}
\bigg\{\bigg[\beta-\frac{1}{2}(\varsigma+\eta)+2\gamma(1+\gamma)+\eta\frac{(\dot{\mathbf{x}}\cdot\mathbf{r}_{\odot})^{2}}{c^{2}r^{2}_{\odot}}\bigg]r^{i}_{\odot}\nonumber\\
&&+\bigg[2(1-\beta)+\varsigma-2\gamma^{2}\bigg]\frac{(\dot{\mathbf{x}}\cdot\mathbf{r}_{\odot})}{c^{2}}\dot{x}^{i}\bigg\}.
\end{eqnarray}
When we consider a case of stationary gravitational field for one
body and $\varsigma=\eta$, our result Eq. (\ref{lightequation}) will
be identical with Ref.~\refcite{kz10}. When we take GR into our model ($\gamma=\beta=\varsigma=\eta=1$), Eq.
(\ref{lightequation}) will reduce to the result in Ref.~\refcite{kk92}.

\subsection{Trajectory of the light-ray in the 2PPN Framework}

The trajectory of the light-ray in the 2PPN framework can be obtained through
integrating Eq. (\ref{lightequation}). However, it is difficult to
derive it directly. So we adopt an iterative method used by Refs.~\refcite{bru91,kk92,kz10}.

But before that, we should find a way to describe the bodies' motion
analytically in the calculation. Chebyshev polynomial could
perfectly deal with the motion of the bodies by interpolated
ephemerides. This has been guaranteed at DE/LE \cite{fwb09}, INPOP \cite{fml14}
and EPM \cite{pp14} ephemeris , due to its
convenience for numerical calculation. For analytical calculation,
however, it is impractical. As an analytical ephemeris, VSOP \cite{sff13}
 used an iterative method by perturbations up to the
the eighth order of masses for the planets in order to
deal with the motion of the bodies. But this approximate method is
still lengthy for our investigation.

A Taylor expansion for the trajectory $\mathbf{x}_{A}$ of the body A
at some moment $t_{A}$ is the simplest and practicable for our
purpose. However, it leads to two sticky issues. One is the order of
the Taylor expansion. The other is the determination of the moment
$t_{A}$. For the first issue, we suppose that the motion of the
bodies, including the Sun, is rectilinear and uniform, that is,
\begin{equation}
\label{bodymotion}
\mathbf{x}_{A}(t)=\mathbf{x}_{A}(t_{A})+\mathbf{v}_{A}(t_{A})(t-t_{A})+\mathcal{O}[(t-t_{A})^{2}].
\end{equation}
Klioner \& Kopeikin \cite{kk92} have pointed out that the residual terms of Eq.
(\ref{bodymotion}) are negligible for the accuracy of $1$ $\mu$as
through a reasonable choice of the moment $t_{A}$. This assumption,
which means the path of a celestial body is approximated to a
straight line, is enough for some measurements conducted in a short
time span since the time of propagation of electromagnetic waves is
very short with respect to the orbital period of a body in the Solar System. For
instance, the time the light takes from the Sun to the Earth is
about 8 minutes while the orbital period of the Earth is about 365
days.

For the second issue, Klioner \& Kopeikin \cite{kk92} indicated that $t_{A}$ can be used to minimize the error in the solution of the light propagation equations caused by the higher-order terms neglected in Eq. (\ref{bodymotion}). Klioner \& Kopeikin \cite{kk92} also showed that the
minimization procedure makes $t_{A}$ equal to the moment of the closest approach of the unperturbed light ray to the body deflecting the light ray. Kopeikin \& Sch\"{a}fer \cite{ks99} proved rigorously by solving Einstein and light ray propagation
equations that at higher astrometric precision $t_{A}$ must be taken as the retarded instant of time corresponding to the retarded
(Li\'{e}nard-Wiechert) solution of linearized gravitational field equations.
Klioner \& Peip \cite{kp03} used the
numerical simulations and showed that it was sufficient to use the
well-known solution for the light propagation in the field of a
motionless mass monopole for the accuracy of $\sim0.2$ $\mu$as and
substituted in that solution the position of the body at the moment
of closest approach. Klioner \& Peip \cite{kp03} found the post-Newtonian
analytical solution for the body being at rest at its position at
the moment of closest approach or at the retarded moment of time are
virtually indistinguishable from each other for the Solar System
applications and showed it attained an accuracy of $\sim0.18$
$\mu$as if we took the form of uniformly moving bodies like Eq.
(\ref{bodymotion}). Therefore, we implement integration technique
like Klioner \& Kopeikin \cite{kk92} to derive the trajectory of light-ray.

We assume the unperturbed light-ray as follows
\begin{equation}
\mathbf{x}_{N}=\mathbf{x}_{0}+c(t-t_{0})\mathbf{\hat{n}},
\end{equation}
where $\mathbf{\hat{n}}$ is a unit vector representing the light direction at $t=-\infty$, $t_{0}$ is an instant on the light path and $\mathbf{x}_{0}$ is the position of photon at $t_{0}$. The photon's coordinates can be written as sum of
perturbations with respect to $\mathbf{x}_{N}$:
\begin{equation}
\mathbf{x}(t)=\mathbf{x}_{N}+\delta\mathbf{x}\equiv\mathbf{x}_{N}
+\delta\mathbf{x}_{1PN}+\delta\mathbf{x}_{Q}+\delta\mathbf{x}_{S}+\delta\mathbf{x}_{2PN}.
\end{equation}
For $\mathcal{F}^{i}_{1PN}$, we use the following assumption for
motion of the bodies
\begin{equation}
\mathbf{r}_{A}(t)=\mathbf{x}(t)-\mathbf{x}_{A}(t)=\mathbf{x}(t)-\mathbf{x}_{A}(t_{A})-\mathbf{v}_{A}(t_{A})(t-t_{A}),
\end{equation}
where $t_{A}$ is the moment of the closest approach between the body
$A$ and the unperturbed light ray. Due to the smallness of
$\mathcal{F}^{i}_{Q}$, $\mathcal{F}^{i}_{S}$ and
$\mathcal{F}^{i}_{2PN}$, in comparison with $\mathcal{F}^{i}_{1PN}$,
it is sufficient to suppose that
\begin{equation}
\mathbf{r}_{A}(t)=\mathbf{x}_{N}(t)-\mathbf{x}_{A}(t_{A}).
\end{equation}
After these, we can obtain the following results
\begin{eqnarray}
\label{lightresult1}
\frac{1}{c}\dot{\mathbf{x}}(t)&=&\mathbf{\hat{n}}
+\frac{1}{c}\delta\dot{\mathbf{x}}_{1PN}(\mathbf{x}_{N}+\delta\mathbf{x}_{1PN})
+\frac{1}{c}\delta\dot{\mathbf{x}}_{Q}(\mathbf{x}_{N})
+\frac{1}{c}\delta\dot{\mathbf{x}}_{S}(\mathbf{x}_{N})
+\frac{1}{c}\delta\dot{\mathbf{x}}_{2PN}(\mathbf{x}_{N}),\\
\label{liahtresult2}
\mathbf{x}(t)&=&\mathbf{x}_{N}(t)+\bigg[\delta\mathbf{x}_{1PN}(\mathbf{x}_{N}+\delta\mathbf{x}_{1PN})-\delta\mathbf{x}_{1PN}(\mathbf{x}_{0})\bigg]
+\bigg[\delta\mathbf{x}_{Q}(\mathbf{x}_{N})-\delta\mathbf{x}_{Q}(\mathbf{x}_{0})\bigg]\nonumber\\
&&+\bigg[\delta\mathbf{x}_{S}(\mathbf{x}_{N})-\delta\mathbf{x}_{S}(\mathbf{x}_{0})\bigg]
+\bigg[\delta\mathbf{x}_{2PN}(\mathbf{x}_{N})-\delta\mathbf{x}_{2PN}(\mathbf{x}_{0})\bigg],
\end{eqnarray}
It is worthy of note that the 2PN terms in our solution actually have two sources, direct
and indirect. The direct part comes from the 2PN order itself. The indirect part comes
from the 1PN terms when the 1PN solution is iterated into itself in order to attain a 2PN
accuracy, namely, we substitute $\mathbf{x}_{N}+\delta\mathbf{x}_{1PN}$ into the trajectory of the light-ray in the 1PN
approximation. During the procedure of iteration, we keep only the 2PN terms related to a
motionless Sun. We can obtain that
\begin{eqnarray}
\frac{1}{c}\delta\dot{\mathbf{x}}_{1PN}(\mathbf{x})&=&-(1+\gamma)\sum_{A}\frac{Gm_{A}}{c^{2}r_{A}}k\bigg\{\frac{\mathbf{\hat{n}}\times(\mathbf{r}_{A}\times\mathbf{k})}{kr_{A}-\mathbf{k}\cdot\mathbf{r}_{A}}
+\mathbf{k}\bigg\},\\
\frac{1}{c}\delta\dot{\mathbf{x}}_{Q}(\mathbf{x})&=&(1+\gamma)\sum_{A}\frac{G}{2c^{2}r^{3}_{A}}\bigg\{
\frac{r_{A}(2r_{A}-\mathbf{\hat{n}}\cdot\mathbf{r}_{A})}{(r_{A}-\mathbf{\hat{n}}\cdot\mathbf{r}_{A})^{2}}\bm{\mathfrak{p}}_{A}
+\bigg[1-3\frac{(\mathbf{\hat{n}}\cdot\mathbf{r}_{A})^{2}}{r^{2}_{A}}\bigg]\bm{\mathfrak{q}}_{A}\nonumber\\
&&-3\frac{\mathbf{\hat{n}}\cdot\mathbf{r}_{A}}{r_{A}}\bm{\mathfrak{r}}_{A}-\bm{\mathfrak{s}}_{A}
\bigg\},\\
\frac{1}{c}\delta\dot{\mathbf{x}}_{S}(\mathbf{x})&=&(1+\gamma)\sum_{A}\frac{G}{c^{3}}\bigg\{
\frac{\mathbf{S}_{A}\times\mathbf{d}_{A}}{r^{3}_{A}}
+\frac{d^{2}_{A}(2r_{A}-\mathbf{\hat{n}}\cdot\mathbf{r}_{A})}{r^{3}_{A}(r_{A}-\mathbf{\hat{n}}\cdot\mathbf{r}_{A})^{2}}
\bigg[(\mathbf{S}_{A}\times\mathbf{\hat{n}})-\frac{\mathbf{d}_{A}}{d^{2}_{A}}\bigg(\mathbf{S}_{A}\cdot(\mathbf{\hat{n}}\times\mathbf{d}_{A})\bigg)\bigg]\nonumber\\
&&-\frac{(\mathbf{S}_{A}\times\mathbf{\hat{n}})}{r_{A}(r_{A}-\mathbf{\hat{n}}\cdot\mathbf{r}_{A})}\bigg\},\\
\frac{1}{c}\delta\dot{\mathbf{x}}_{2PN}(\mathbf{x})&=&-\frac{1}{2}\eta\frac{G^{2}m^{2}_{\odot}}{c^{4}r^{4}_{\odot}}(\mathbf{\hat{n}}\cdot\mathbf{r}_{\odot})\mathbf{r}_{\odot}
+\frac{G^{2}m^{2}_{\odot}}{c^{4}}\mathbf{d}_{\odot}\bigg\{(1+\gamma)^{2}\frac{1}{r_{\odot}(r_{\odot}-\mathbf{\hat{n}}\cdot\mathbf{r}_{\odot})}
\bigg(\frac{2}{r_{\odot}}
+\frac{1}{r_{\odot}-\mathbf{\hat{n}}\cdot\mathbf{r}_{\odot}}\bigg)\nonumber\\
&&+[\beta-\frac{1}{4}(2\varsigma+\eta)-2(1+\gamma)]\frac{1}{d^{2}_{\odot}}\bigg[
\frac{\mathbf{\hat{n}}\cdot\mathbf{r}_{\odot}}{r^{2}_{\odot}}
+\frac{1}{d_{\odot}}\bigg(\frac{\pi}{2}+\arctan\frac{\mathbf{\hat{n}}\cdot\mathbf{r}_{\odot}}{d_{\odot}}\bigg)\bigg]\bigg\}\nonumber\\
&&+\frac{G^{2}m^{2}_{\odot}}{c^{4}r_{\odot}}\mathbf{\hat{n}}\bigg[\bigg(2\gamma(1+\gamma)+\beta-\frac{1}{2}\varsigma\bigg)\frac{1}{r_{\odot}}
-(1+\gamma)^{2}\frac{1}{r_{\odot}-\mathbf{\hat{n}}\cdot\mathbf{r}_{\odot}}\bigg],
\end{eqnarray}
and
\begin{eqnarray}
\delta\mathbf{x}_{1PN}(\mathbf{x})&=&-(1+\gamma)\sum_{A}\frac{Gm_{A}}{c^{2}}\bigg\{\frac{\mathbf{\hat{n}}\times(\mathbf{r}_{A}\times\mathbf{k})}{kr_{A}-\mathbf{k}\cdot\mathbf{r}_{A}}
+\mathbf{k}\ln\bigg(kr_{A}+\mathbf{k}\cdot\mathbf{r}_{A}\bigg)
\bigg]\bigg\},\\
\delta\mathbf{x}_{Q}(\mathbf{x})&=&(1+\gamma)\sum_{A}\frac{G}{2c^{2}d^{2}_{A}}\bigg\{\frac{r_{A}+\mathbf{\hat{n}}\cdot\mathbf{r}_{A}}{r_{A}-\mathbf{\hat{n}}\cdot\mathbf{r}_{A}}\bm{\mathfrak{p}}_{A}
+\frac{d^{2}_{A}(\mathbf{\hat{n}}\cdot\mathbf{r}_{A})}{r^{3}_{A}}\bm{\mathfrak{q}}_{A}
+\frac{d^{2}_{A}}{r^{2}_{A}}\bm{\mathfrak{r}}_{A}-\frac{\mathbf{\hat{n}}\cdot\mathbf{r}_{A}}{r_{A}}\bm{\mathfrak{s}}_{A}
\bigg\},\\
\delta\mathbf{x}_{S}(\mathbf{x})&=&(1+\gamma)\sum_{A}\frac{G}{c^{3}r_{A}}\bigg\{\frac{\mathbf{\hat{n}}\cdot\mathbf{r}_{A}}{d^{2}_{A}}(\mathbf{S}_{A}\times\mathbf{d}_{A})
-\frac{\mathbf{d}_{A}(r_{A}+\mathbf{\hat{n}}\cdot\mathbf{r}_{A})}{d^{2}_{A}(r_{A}-\mathbf{\hat{n}}\cdot\mathbf{r}_{A})}[\mathbf{S}_{A}\cdot(\mathbf{\hat{n}}\times\mathbf{d}_{A})]\nonumber\\
&&+\frac{\mathbf{\hat{n}}\cdot\mathbf{r}_{A}}{r_{A}-\mathbf{\hat{n}}\cdot\mathbf{r}_{A}}(\mathbf{S}_{A}\times\mathbf{\hat{n}})\bigg\},\\
\delta\mathbf{x}_{2PN}(\mathbf{x})&=&\frac{1}{4}\eta\frac{G^{2}m^{2}_{\odot}}{c^{4}r^{2}_{\odot}}\mathbf{r}_{\odot}
+\frac{G^{2}m^{2}_{\odot}}{c^{4}}\mathbf{\hat{n}}\bigg[\bigg(\beta-\frac{1}{4}(2\varsigma+\eta)-2(1+\gamma)\bigg)
\frac{1}{d_{\odot}}\arctan\frac{\mathbf{\hat{n}}\cdot\mathbf{r}_{\odot}}{d_{\odot}}\nonumber\\
&&-(1+\gamma)^{2}\frac{1}{r_{\odot}-\mathbf{\hat{n}}\cdot\mathbf{r}_{\odot}}\bigg]
+\frac{G^{2}m^{2}_{\odot}}{c^{4}}\mathbf{d}_{\odot}\bigg\{(1+\gamma)^{2}\frac{1}{(r_{\odot}
-\mathbf{\hat{n}}\cdot\mathbf{r}_{\odot})^{2}}\nonumber\\
&&+[\beta-\frac{1}{4}(2\varsigma+\eta)-2(1+\gamma)]
\frac{\mathbf{\hat{n}}\cdot\mathbf{r}_{\odot}}{d^{3}_{\odot}}
\bigg[\frac{\pi}{2}
+\arctan\bigg(\frac{\mathbf{\hat{n}}\cdot\mathbf{r}_{\odot}}{d_{\odot}}\bigg)\bigg]
\bigg\},
\end{eqnarray}
where
$\mathbf{d}_{A}=\mathbf{\hat{n}}\times(\mathbf{r}_{A}\times\mathbf{\hat{n}})$
is an impact parameter for the body $A$ which is the closest
approach of the unperturbed light ray, $d_{A}=|\mathbf{d}_{A}|$,
$\mathbf{r}_{A}=\mathbf{x}_{N}-\mathbf{x}_{A}(t)$,
$r_{A}=|\mathbf{r}_{A}|$,
$\mathbf{k}=\mathbf{\hat{n}}-\mathbf{v}_{A}(t_{A})/c$,
$k=|\mathbf{k}|$ and
\begin{eqnarray}
\mathfrak{p}^{i}_{A}&=&2J^{<ik>}_{A}\frac{d^{k}_{A}}{r_{A}}-2J^{<jk>}_{A}\frac{d^{j}_{A}}{r_{A}}\hat{n}^{k}\hat{n}^{i}
-J^{<jk>}_{A}\bigg(\hat{n}^{j}\hat{n}^{k}+\frac{4d^{j}_{A}d^{k}_{A}}{d^{2}_{A}}\bigg)\frac{d^{i}_{A}}{r_{A}},\\
\mathfrak{q}^{i}_{A}&=&2J^{<jk>}_{A}\frac{\hat{n}^{j}d^{k}_{A}}{d^{2}_{A}}d^{i}_{A}+J^{<jk>}_{A}\bigg(\hat{n}^{j}\hat{n}^{k}-\frac{d^{j}_{A}d^{k}_{A}}{d^{2}_{A}}\bigg)\hat{n}^{i},\\
\mathfrak{r}^{i}_{A}&=&2J^{<jk>}_{A}\frac{d^{j}_{A}}{r_{A}}\hat{n}^{k}\hat{n}^{i}-J^{<jk>}_{A}\bigg(\hat{n}^{j}\hat{n}^{k}
-\frac{d^{j}_{A}d^{k}_{A}}{d^{2}_{A}}\bigg)\frac{d_{A}^{i}}{r_{A}},\\
\mathfrak{s}^{i}_{A}&=&2J^{<ik>}_{A}\hat{n}^{k}
-4J^{<jk>}_{A}\frac{\hat{n}^{j}d^{k}_{A}}{d^{2}_{A}}d^{i}_{A}
-J^{<jk>}_{A}\bigg(\hat{n}^{j}\hat{n}^{k}-2\frac{d^{j}_{A}d^{k}_{A}}{d^{2}_{A}}\bigg)\hat{n}^{i}.
\end{eqnarray}

If we only consider a static case for one body and assume $\eta=\varsigma$, Eqs.
(\ref{lightresult1}) and (\ref{liahtresult2}) will return to the
results of Ref.~\refcite{kz10}. When we neglect $\mathcal{F}^{i}_{2PN}$ and $\mathcal{F}^{i}_{S}$,
Eqs. (\ref{lightresult1}) and (\ref{liahtresult2}) will go back to
the results of Ref.~\refcite{kli03}. When we consider GR
($\gamma=\beta=\varsigma=\eta=1$), Eqs. (\ref{lightresult1}) and
(\ref{liahtresult2}) will reduce to the results of Ref.~\refcite{kk92}.

\section{Angular measurement}

\subsection{Construction of a gauge-invariant angle in the 2PPN framework}
In practical astronomic measurements, a differential measurement
is a more powerful method. This concept is employed by LATOR mission
\cite{tsn04} through a skinny triangle formed by two
spacecrafts and the ISS. In what follows, we mainly focus on
discussing LATOR-like missions. Firstly, we construct a gauge-invariant
angle $\theta$ between the directions of the two incoming photons
based on Ref.~\refcite{bru91}. It reads
\begin{equation}
\label{cosdef}
\cos\theta=\frac{h_{\alpha\beta}K^{\alpha}_{1}K^{\beta}_{2}}
{\sqrt{h_{\alpha\beta}K^{\alpha}_{1}K^{\beta}_{1}}\sqrt{h_{\alpha\beta}K^{\alpha}_{2}K^{\beta}_{2}}},
\end{equation}
where the spatial projection operator is
\begin{equation}
h_{\alpha\beta}=g_{\alpha\beta}+u_{\alpha}u_{\beta},
\end{equation}
which projects the two incoming photons onto the hypersurface
orthogonal to the observer's four-velocity $u^{\alpha}\equiv
\mathrm{d}x^{\alpha}/c\mathrm{d}\tau$ and $K^{\alpha}_{1}\equiv
\mathrm{d}x^{\alpha}_{1}(t)/\mathrm{d}t$, and $K^{\beta}_{2}\equiv
\mathrm{d}x^{\beta}_{2}(t)/\mathrm{d}t$ are the tangent vectors
of the paths $x^{\alpha}_{1}(t)$ and $x^{\beta}_{2}(t)$ of the two
incoming photons. Then, we obtain
\begin{eqnarray}
\label{cos}
\cos\theta&=&(\mathbf{\hat{n}}_{1}\cdot\mathbf{\hat{n}}_{2})
+\overset{(1)}{f}_{obs}+\bigg(\overset{(2)}{f}_{obs}+\overset{(2)}{f}_{1PN}+\overset{(2)}{f}_{Q}\bigg)
+\bigg(\overset{(3)}{f}_{obs}+\overset{(3)}{f}_{obs\times
1PN}+\overset{(3)}{f}_{S}\bigg)\nonumber\\
&&+\bigg(\overset{(4)}{f}_{obs}+\overset{(4)}{f}_{obs\times1PN}+\overset{(4)}{f}_{2PN}\bigg),
\end{eqnarray}
where the first term is the angle between the unperturbed paths of
the photons from two signals, $\overset{(n)}{f}$ denotes terms of
order $\epsilon^{n}$, the subscript ``$obs$" denotes terms
related to the observer's velocity, subscript ``$1PN$" denotes the contribution from the monopoles
and orbital motions of gravitating bodies, subscript ``$Q$"
denotes the terms from their
quadrupole moments, subscript
``$obs\times 1PN$" denotes the coupling terms of the bodies' mass
and observer's velocity, subscript ``$S$" denotes the terms of bodies' spin and subscript ``$\odot$" denotes the Sun monopole contribution.
The expressions for these functions are presented in
\ref{appendixC}. Obviously, the position of the photon at
the moment $t$ of observation coincides with the position of the
spacecraft (observer) so that
\begin{eqnarray}
\label{constrant}
\mathbf{x}_{obs}=\mathbf{x}_{01}+c(t-t_{01})\mathbf{\hat{n}}_{1}+\delta\mathbf{x}_{1}
=\mathbf{x}_{02}+c(t-t_{02})\mathbf{\hat{n}}_{2}+\delta
\mathbf{x}_{2},
\end{eqnarray}
where $(t_{01},\mathbf{x}_{01})$ denotes the moment and position of the light signal 1 of emission
and $(t_{02},\mathbf{x}_{02})$ for
the light signal $2$ respectively. Eq. (\ref{constrant}) also give a constraint at the
moment of measurement that
\begin{equation}
\mathbf{r}_{1A}=\mathbf{r}_{2A}=\mathbf{r}_{obsA}\equiv\mathbf{r}_{A}.
\end{equation}

Eq. (\ref{cos}) can be expanded with respect to the parameter
$\epsilon$ as
\begin{eqnarray}
\label{theta}
\theta(t)&=&\vartheta_{0}+\overset{(1)}{\vartheta}_{obs}+\bigg(\overset{(2)}{\vartheta}_{obs}
+\overset{(2)}{\vartheta}_{1PN}+\overset{(2)}{\vartheta}_{Q}\bigg)
+\bigg(\overset{(3)}{\vartheta}_{obs} +\overset{(3)}{\vartheta}_{OM}
+\overset{(3)}{\vartheta}_{S}\bigg)\nonumber\\
&&+\bigg(\overset{(4)}{\vartheta}_{obs}
+\overset{(4)}{\vartheta}_{2PN}\bigg),
\end{eqnarray}
where $\vartheta_{0}\in(0,\pi)$ is the angle between the unperturbed
light paths from two given sources
\begin{equation}
\vartheta_{0}=\arccos(\mathbf{\hat{n}}_{1}\cdot\mathbf{\hat{n}}_{2}),
\end{equation}
and $\overset{(n)}{\vartheta}_{obs}$ is the deflection
angle due to the observer's motion in terms of order $\epsilon^{n}$,
$\overset{(2)}{\vartheta}_{1PN}$ is the 1PN deflection
angle due to the spherically symmetric field of gravitating bodies,
$\overset{(2)}{\vartheta}_{Q}$ is
due to quadrupole moment,
$\overset{(3)}{\vartheta}_{OM}$ is
due to the orbital motions of the bodies,
$\overset{(3)}{\vartheta}_{S}$ is
due to the spin of the bodies and
$\epsilon^{4}\overset{(4)}{\vartheta}_{2PN}$ is the 2PN deflection
angle due to the spherically symmetric field of the Sun (see
\ref{appendixD} for detailed expressions).

\subsection{Special cases}

As one kind of the verification, our results could reduce to some
special cases and some
known results, such as
aberration of special relativity, 1PN light deflection, 2PN light
deflection and Light deflection due to spins.

\subsubsection{Aberration}

In Eq. (\ref{theta}), we pick
up the terms due to the velocity of the observer
\begin{eqnarray}
\theta&=&\vartheta_{0}+\overset{(1)}{\vartheta}_{obs}+\overset{(2)}{\vartheta}_{obs},
\end{eqnarray}
and assume that one signal propagates along the direction of motion of the observer, namely,
$\mathbf{v}_{obs}=v\mathbf{\hat{n}}_{1}$. Then, we obtain
\begin{equation}
\theta=\vartheta_{0}+\epsilon
v\sin\vartheta_{0}+\epsilon^{2}\frac{1}{2}v^{2}\sin\vartheta_{0}\cos\vartheta_{0}.
\end{equation}
This is just the aberration given by special relativity after ignoring $\mathcal{O}(\epsilon^{3})$.

\subsubsection{1PN light deflection}

The path of light is bent in a gravitational
field, which represents by
\begin{eqnarray}
\theta&=&\overset{(2)}{\vartheta}_{1PN}.
\end{eqnarray}
We assume that only the Sun is considered and one of the two light rays is just along the
line connecting the Sun and the observer and without any bending, namely source 2. Then,
we obtain
\begin{eqnarray}
\label{deflection}
\overset{(2)}{\vartheta}_{1PN}&\approx&\bigg(\frac{1+\gamma}{2}\bigg)\frac{4Gm_{\odot}}{c^{2}d_{1\odot}}\bigg(\frac{1+\cos\vartheta_{0}}{2}\bigg),
\end{eqnarray}
where we use
$\mathbf{\hat{n}}_{2}\cdot\mathbf{d}_{1A}/d_{1A}=\sin\vartheta_{0}$,
$\mathbf{\hat{n}}_{1}\cdot\mathbf{r}_{A}/r_{A}\simeq\cos\vartheta_{0}$
and
$d^2_{1A}=r^{2}_{A}-(\mathbf{\hat{n}}_{1}\cdot\mathbf{r}_{A})^{2}$.
For GR ($\gamma=1$), When $\cos\vartheta_{0}=1$, our results
reduces to the famous infinity-infinity light deflection formula.

Brumberg\cite{bru91} discussed the angle of the two incoming photons in 1PN
general relativity for a $N$-body system. Now, we discuss our result about
1PN gravitational deflection, namely,
\begin{eqnarray}
\cos\theta&=&(\mathbf{\hat{n}}_{1}\cdot\mathbf{\hat{n}}_{2})+\overset{(2)}{f}_{1PN}\nonumber\\
&=&(\mathbf{\hat{n}}_{1}\cdot\mathbf{\hat{n}}_{2})-\epsilon^{2}(1+\gamma)\sum_{A}\frac{Gm_{A}}{r_{A}}
\bigg\{\frac{\mathbf{\hat{n}}_{1}\cdot[\mathbf{\hat{n}}_{2}\times(\mathbf{r}_{A}\times\mathbf{\hat{n}}_{2})]}{r_{A}-\mathbf{\hat{n}}_{2}\cdot\mathbf{r}_{A}}
+\frac{\mathbf{\hat{n}}_{2}\cdot[\mathbf{\hat{n}}_{1}\times(\mathbf{r}_{A}\times\mathbf{\hat{n}}_{1})]}{r_{A}-\mathbf{\hat{n}}_{1}\cdot\mathbf{r}_{A}}\bigg\}\nonumber\\
&=&(\mathbf{\hat{n}}_{1}\cdot\mathbf{\hat{n}}_{2})+\epsilon^{2}(1+\gamma)\sum_{A}\frac{Gm_{A}}{r_{A}}\bigg\{
\frac{\mathbf{r}_{A}\times\mathbf{\hat{n}}_{1}}{r_{A}-\mathbf{\hat{n}}_{1}\cdot\mathbf{r}_{A}}
-\frac{\mathbf{r}_{A}\times\mathbf{\hat{n}}_{2}}{r_{A}-\mathbf{\hat{n}}_{2}\cdot\mathbf{r}_{A}}\bigg\}\cdot(\mathbf{\hat{n}}_{1}\times\mathbf{\hat{n}}_{2}).
\end{eqnarray}
When $\gamma=1$, this is just the result of Ref.~\refcite{bru91}.

\subsubsection{2PN light deflection}

By extending the previous calculation to 2PN order, we have
$\arctan(\hat{\mathbf{n}}_{1}\cdot\mathbf{r}_{\odot}/d_{1\odot})\approx\pi/2$,
$d_{1\odot}/r_{\odot}\approx0$ and
\begin{eqnarray}
\overset{(4)}{\vartheta}_{2PN}&\approx&\bigg\{-2(1+\gamma)^{2}+[2(1+\gamma)-\beta+\frac{1}{4}(2\varsigma+\eta)]\pi\bigg\}
\frac{G^{2}m_{\odot}^{2}}{c^{4}d^{2}_{1\odot}}.
\end{eqnarray}
If the above result returns to GR ($\gamma=\beta=\varsigma=\eta=1$),
it coincides with the result of GR within harmonic gauge
. If $(2\varsigma+\eta)\equiv3\Lambda$, it will coincide with the
result of Ref.~\refcite{rm82}. In the case of $\varsigma=\eta$, it
will reproduce the result of Ref.~\refcite{kz10}.

\subsubsection{Light deflection due to spins}

The spin of a body can also produce a light deflection. It belongs
to one part of gravitomagnetic field. It reads as
\begin{eqnarray}
\theta&=&\overset{(3)}{\vartheta}_{S}.
\end{eqnarray}
Repeating previous calculations, we obtain
\begin{eqnarray}
\theta&\approx&2(1+\gamma)\frac{|\mathbf{S}_{A}|}{c^{3}d^{2}_{1A}}
+4(1+\gamma)\frac{[\mathbf{S}_{A}\cdot(\mathbf{\hat{n}}_{1}\times\mathbf{d}_{1A})]}{c^{3}d^{3}_{1A}}.
\end{eqnarray}
This is just the result of Ref.~\refcite{km02} with $\gamma=1$ with
turning their vector angle for deflection by reason of spin to a
scalar quantity.

\subsection{A MODEL FOR LATOR-LIKE MISSIONS}

In this section, we apply our model to LATOR-like missions by qualitative estimate.We assume there are three spacecrafts in an orbit circling the Sun at the same distance as
the Earth. Two of them carry a light signal emitter on board and the third one carries
a light signal receiver. When the two spacecrafts and the third one are on the opposite
side of the Sun, the experiment will be conducted. When a light signal 1 emitted by one
of two spacecrafts passes by the limb of the Sun, another signal 2 is emitted by the other
spacecraft which has quite a distance from the Sun. The receiver located on the third
spacecraft which is on the opposite side of those two spacecrafts measures the angle between
the two incoming photons. In the case of LATOR
\cite{tsn04}, the observer will be set on the ISS.

\begin{table}[ph]
\tbl{The estimation of the terms in Eq. (\ref{practical})}
{\begin{tabular}{@{}cccccccccccc@{}} \toprule
Term1 & Term2 & Term3 &Term4 &  Term5 & Term6 &Term7 & Term8 &Term9 &  Term10 &  Term11 & Term12 \\
\colrule
$3.14$mas & $18.02\mu$as & $1''.75$ & $0''.47$ &  $83.06\mu$as & $492.76\mu$as &
$4.13\mu$as & $1.53\mu$as & $3.05\mu$as &  $7.44\mu$as &  $11.32\mu$as & $1.20\mu$as \\\colrule
\end{tabular} \label{practical1}}
\end{table}

We assume that the velocity of the observer in SSBRS is near the orbital velocity of the
Earth. In this case, we need to
consider three bodies' gravitational fields: the Sun, Mercury and
Venus. We can neglect the effect of gravitational field of the Earth through making the
position of the observer away from the Earth. However, for the case of ISS, the gravitational
field of the Earth affects the motion of ISS and this effect can be represented by equations
of motion for ISS. The influence of other gravitating bodies in the
Solar System on the light propagation in this case can be neglected.
For LATOR, signals are emitted inside the Solar System, but we could extended the trajectories to
$t=-\infty$ to make make $\hat{\mathbf{n}}_{1}$ and $\hat{\mathbf{n}}_{2}$
meaningful.
With the angle between the initial emitting directions of two light
signals 1 and 2
\begin{equation}
\vartheta_{0}=\mathbf{\hat{n}}_{1}\cdot\mathbf{\hat{n}}_{2}\approx1^{\circ},
\end{equation}
it means that the distance between these two spacecrafts is about
$5.22\times10^{9}$ m. We can simplify Eq. (\ref{theta}) for this
practical case as follows
\begin{eqnarray}
\label{practical} \theta&=&\vartheta_{0}
+\underbrace{\epsilon(\mathbf{\hat{n}}_{2}\cdot\mathbf{v}_{obs})\tan\bigg(\frac{\vartheta_{0}}{2}\bigg)}_{Term1}
-\underbrace{\epsilon^{2}v^{2}_{obs}\tan\bigg(\frac{\vartheta_{0}}{2}\bigg)}_{Term2}
+\underbrace{\epsilon^{2}2(1+\gamma)\frac{Gm_{\odot}}{d_{1\odot}}}_{Term3}\nonumber\\
&&+\underbrace{\epsilon^{2}2(1+\gamma)\frac{Gm_{\odot}}{d_{2\odot}}}_{Term4}
+\underbrace{\epsilon^{2}2(1+\gamma)\frac{Gm_{Mer}}{d_{1Mer}}}_{Term5}
+\underbrace{\epsilon^{2}2(1+\gamma)\frac{Gm_{V}}{d_{1V}}}_{Term6}\nonumber\\
&&+\underbrace{\epsilon^{2}2(1+\gamma)\frac{Gm_{V}}{d_{2V}}}_{Term7}
-\underbrace{\epsilon^{3}2(1+\gamma)\tan\bigg(\frac{\vartheta_{0}}{2}\bigg)\frac{Gm_{\odot}}{d_{1\odot}}\frac{\mathbf{v}_{obs}\cdot\mathbf{d}_{1\odot}}{d_{1\odot}}}_{Term8}\nonumber\\
&&+\underbrace{\epsilon^{3}2(1+\gamma)\frac{Gm_{\odot}}{d_{1\odot}}(\mathbf{\hat{n}}_{2}\cdot\mathbf{v}_{obs})}_{Term9}
-\underbrace{\epsilon^{4}2(1+\gamma)^{2}\frac{G^{2}m^{2}_{\odot}}{d^{2}_{1\odot}}}_{Term10}
\nonumber\\
&&-\underbrace{\epsilon^{4}[\beta-\frac{1}{4}(2\varsigma+\eta)-2(1+\gamma)^{2}]\frac{G^{2}m^{2}_{\odot}}{d^{2}_{1\odot}}
\bigg[\frac{\pi}{2}+\arctan\bigg(\frac{\mathbf{\hat{n}}_{1}\cdot\mathbf{r}_{\odot}}{d_{1\odot}}\bigg)\bigg]}_{Term11}\nonumber\\
&&-\underbrace{\epsilon^{4}[\beta-\frac{1}{4}(2\varsigma+\eta)-2(1+\gamma)^{2}]\frac{G^{2}m^{2}_{\odot}}{d^{2}_{2\odot}}
\bigg[\frac{\pi}{2}+\arctan\bigg(\frac{\mathbf{\hat{n}}_{2}\cdot\mathbf{r}_{\odot}}{d_{2\odot}}\bigg)\bigg]}_{Term12}\nonumber\\
&&+\mathcal{O}(<1\mu as),
\end{eqnarray}
with the cut-off precision of $\sim1$ $\mu$as. Here we consider the
largest influence of Mercury (subscript ``Mer") and Venus (subscript
``V") on the measurement when the light signals might pass by the
limbs of them. We estimate the order of these terms on Eq.
(\ref{practical}) which are listed on Table \ref{practical1}.

For LATOR mission, it will attain the precision of $0.01$ $\mu$as.
It needs to
consider a complete form at $c^{-4}$ of $g_{ij}$ and $g_{00}$.
Various terms at $c^{-4}$ of $g_{ij}$ and $g_{00}$ respectively are
$\sum_{A}\frac{Gm_{A}}{c^{4}r_{A}}v^{2}_{A}$,
$\sum_{A}\frac{G^{2}m^{2}_{A}}{c^{4}r^{2}_{A}}$,
$\sum_{A}\sum_{B\neq A}\frac{Gm_{A}m_{B}}{c^{4}r_{A}r_{AB}}$,
$\sum_{A}\sum_{B\neq A}\frac{Gm_{A}m_{B}}{c^{4}r_{A}r_{B}}$,
$\sum_{A}\frac{Gm_{A}}{c^{4}r_{A}}v^{i}_{A}v^{j}_{A}$,
$\sum_{A}\frac{Gm_{A}}{c^{4}r^{4}_{A}}r^{i}_{A}r^{j}_{A}$,
$\sum_{A}\sum_{B\neq
A}\frac{Gm_{A}m_{B}}{c^{4}r^{3}_{AB}(r_{A}+r_{B}+r_{AB})}r^{i}_{AB}r^{j}_{AB}$
and $\sum_{A}\sum_{B\neq
A}\frac{Gm_{A}m_{B}}{c^{4}(r_{A}+r_{B}+r_{AB})^{2}}
(\frac{r^{i}_{AB}}{r_{AB}}-\frac{r^{i}_{A}}{r_{A}})(\frac{r^{j}_{AB}}{r_{AB}}+\frac{r^{i}_{B}}{r_{B}})$.
These terms have be shown by Ref.~\refcite{dx12}. These terms can be considered as different combinations
between terms in Table \ref{ta2} and in Table \ref{ta4}. If the accuracy of the model is required to reach 0.01 $\mu$as, we need to deal with two problems: (1) integrating the two-body terms in the 2PN order and (2) introducing new parameters of the coupling terms in the 2PN order.
Besides, LATOR also measures laser ranging for three sides
of the triangle, it needs to derive the light time solution. And the spacecrafts in LATOR mission are constantly moving. The triangle formed by the
spacecrafts are also changing. $\vartheta_{0}\approx1^{\circ}$ is just a special case which presents the maximal
effect of this measurement. We will study these issues
numerically in our next move.

\section{CONCLUSIONS AND PROSPECTS}
In this paper, we present light propagation and a gauge-invariant
angular measurement in the 2PPN framework by introducing
two new parameters $\varsigma$ and $\eta$ besides the two PPN
parameters $\gamma$ and $\beta$. In the framework, we consider all kinds of relativistic effects on light propagation in SSBRS, which are monopole and
quadrupole moments of the bodies in the Solar System, their motions and their
gravitomagnetic fields. With the derivation of a
gauge-invariant angle between the directions of two incoming
photons, we further discuss a practical astronomic observation,
namely, an observer on a spacecraft measures the angle between
the two incoming photons emitted separately two spacecrafts which
are all at an orbit circling the Sun at the same distance as the
Earth. Given attaining a level of $\sim1$ $\mu$as for space astrometry missions in the near future, like LATOR
mission, the terms at this level are listed.

In this work, We approximate planetary motions as linear motion. This approximation is not valid for the time scale comparable with the orbital periods of planets. Our next move is to study
this problem numerically by integrating the equations of motion of
$N$-body and light simultaneously. How to parameterize the 2PN coupling terms
in these experiments is another issue that will be investigated.

\section*{Acknowledgments}

The author especially would like to thank Professor Tian-Yi
Huang of Nanjing University for his fruitful discussions. This work is funded by the National Natural Science Foundation
of China (Grant No. 11473072) and the Fundamental Research Program of Jiangsu Province of
China (Grant No. BK20131461).

\appendix

\section{Functions in $\cos\theta$ \label{appendixC}}

The expressions of the functions in Eq. (\ref{cos}) are
\begin{eqnarray}
\overset{(1)}{f}_{obs}&=&\epsilon\sum_{k=1}^{2}(\mathbf{\hat{n}}_{k}\cdot\mathbf{v}_{obs})\bigg[
(\mathbf{\hat{n}}_{1}\cdot\mathbf{\hat{n}}_{2})-1\bigg],
\end{eqnarray}
\begin{eqnarray}
\overset{(2)}{f}_{obs}&=&\epsilon^{2}\bigg[\sum_{k=1}^{2}(\mathbf{\hat{n}}_{k}\cdot\mathbf{v}_{obs})^{2}
+\frac{1}{2}\sum_{k,j=1}^{2}\sum_{k\neq
j}(\mathbf{\hat{n}}_{k}\cdot\mathbf{v}_{obs})(\mathbf{\hat{n}}_{j}\cdot\mathbf{v}_{obs})-v^{2}_{obs}\bigg]\bigg[
(\mathbf{\hat{n}}_{1}\cdot\mathbf{\hat{n}}_{2})-1\bigg],
\end{eqnarray}
\begin{eqnarray}
\overset{(2)}{f}_{1PN}&=&\sum_{h,j=1}^{2}\sum_{h\neq
j}\mathbf{\hat{n}}_{j}\cdot\frac{1}{c}\delta\dot{\mathbf{x}}_{1PN}(\mathbf{x}_{N}+\delta\mathbf{x}_{1PN})|_{h}\nonumber\\
&&-\sum_{j=1}^{2}(\mathbf{\hat{n}}_{1}\cdot\mathbf{\hat{n}}_{2})
\mathbf{\hat{n}}_{j}\cdot\frac{1}{c}\delta\dot{\mathbf{x}}_{1PN}(\mathbf{x}_{N}+\delta\mathbf{x}_{1PN})|_{j},
\end{eqnarray}
\begin{eqnarray}
\overset{(2)}{f}_{Q}&=&\sum_{h,j=1}^{2}\sum_{h\neq
j}\mathbf{\hat{n}}_{j}\cdot\frac{1}{c}\delta\dot{\mathbf{x}}_{Q}(\mathbf{x}_{N})|_{h}
-\sum_{j=1}^{2}(\mathbf{\hat{n}}_{1}\cdot\mathbf{\hat{n}}_{2})
\mathbf{\hat{n}}_{j}\cdot\frac{1}{c}\delta\dot{\mathbf{x}}_{Q}(\mathbf{x}_{N})|_{j},
\end{eqnarray}
\begin{eqnarray}
\overset{(3)}{f}_{obs}&=&\epsilon^{3}\bigg[\sum_{k=1}^{2}(\mathbf{\hat{n}}_{k}\cdot\mathbf{v}_{obs})^{3}
-\sum_{k=1}^{2}(\mathbf{\hat{n}}_{k}\cdot\mathbf{v}_{obs})v^{2}_{obs}
+\sum_{k,j=1}^{2}\sum_{k\neq
j}(\mathbf{\hat{n}}_{k}\cdot\mathbf{v}_{obs})^{2}(\mathbf{\hat{n}}_{j}\cdot\mathbf{v}_{obs})\bigg]\nonumber\\
&&\times\bigg[(\mathbf{\hat{n}}_{1}\cdot\mathbf{\hat{n}}_{2})-1\bigg],
\end{eqnarray}
\begin{eqnarray}
\overset{(3)}{f}_{obs\times
1PN}&=&\epsilon\sum_{k=1}^{2}\mathbf{v}_{obs}\cdot\frac{1}{c}\delta\dot{\mathbf{x}}_{1PN}(\mathbf{x}_{N})|_{k}
\bigg[(\mathbf{\hat{n}}_{1}\cdot\mathbf{\hat{n}}_{2})-1\bigg]
+\epsilon\sum_{j,h,l=1}^{2}(\mathbf{\hat{n}}_{l}\cdot\mathbf{v}_{obs})\hat{n}_{j}\cdot\frac{1}{c}\delta\dot{\mathbf{x}}_{1PN}(\mathbf{x}_{N})|_{h}\nonumber\\
&&-3\epsilon\sum_{j=1}^{2}(\mathbf{\hat{n}}_{1}\cdot\mathbf{\hat{n}}_{2})(\mathbf{\hat{n}}_{j}\cdot\mathbf{v}_{obs})
\mathbf{\hat{n}}_{j}\cdot\frac{1}{c}\delta\dot{\mathbf{x}}_{1PN}(\mathbf{x}_{N})|_{j}\nonumber\\
&&-\epsilon\sum_{j,h=1}^{2}\sum_{j\neq
h}(\mathbf{\hat{n}}_{1}\cdot\mathbf{\hat{n}}_{2})(\mathbf{\hat{n}}_{j}\cdot\mathbf{v}_{obs})
\mathbf{\hat{n}}_{h}\cdot\frac{1}{c}\delta\dot{\mathbf{x}}_{1PN}(\mathbf{x}_{N})|_{h},
\end{eqnarray}
\begin{eqnarray}
\overset{(3)}{f}_{S}&=&\sum_{j,h=1}^{2}\sum_{j\neq
h}\mathbf{\hat{n}}_{j}\cdot\frac{1}{c}\delta\dot{\mathbf{x}}_{S}(\mathbf{x}_{N})|_{h}
-\sum_{j=1}^{2}(\mathbf{\hat{n}}_{1}\cdot\mathbf{\hat{n}}_{2})
\mathbf{\hat{n}}_{j}\cdot\frac{1}{c}\delta\dot{\mathbf{x}}_{S}(\mathbf{x}_{N})|_{j},
\end{eqnarray}
\begin{eqnarray}
\overset{(4)}{f}_{obs}&=&\epsilon^{4}\bigg[\sum_{k=1}^{2}(\mathbf{\hat{n}}_{k}\cdot\mathbf{v}_{obs})^{4}
-\sum_{k=1}^{2}(\mathbf{\hat{n}}_{k}\cdot\mathbf{v}_{obs})^{2}v^{2}_{obs}
-\frac{1}{2}\sum_{k,j=1}^{2}\sum_{k\neq
j}(\mathbf{\hat{n}}_{k}\cdot\mathbf{v}_{obs})(\mathbf{\hat{n}}_{j}\cdot\mathbf{v}_{obs})v^{2}_{obs}
\nonumber\\
&&+\sum_{k,j=1}^{2}\sum_{k\neq
j}(\mathbf{\hat{n}}_{k}\cdot\mathbf{v}_{obs})^{3}(\mathbf{\hat{n}}_{j}\cdot\mathbf{v}_{obs})
+\frac{1}{2}\sum_{k,j=1}^{2}\sum_{k\neq
j}(\mathbf{\hat{n}}_{k}\cdot\mathbf{v}_{obs})^{2}(\mathbf{\hat{n}}_{j}\cdot\mathbf{v}_{obs})^{2}\bigg]\bigg[
(\mathbf{\hat{n}}_{1}\cdot\mathbf{\hat{n}}_{2})-1\bigg],
\end{eqnarray}
\begin{eqnarray}
\overset{(4)}{f}_{obs\times1PN}&=&-(1+\gamma)\epsilon^{4}\sum_{k=1}^{2}\frac{Gm_{\odot}}{r_{obs\odot}}(\mathbf{\hat{n}}_{k}\cdot\mathbf{v}_{obs})^{2}(\mathbf{\hat{n}}_{1}\cdot\mathbf{\hat{n}}_{2})
+(1+\gamma)\epsilon^{4}\sum_{k,j=1}^{2}\sum_{k\neq
j}\frac{Gm_{\odot}}{r_{obs\odot}}(\mathbf{\hat{n}}_{k}\cdot\mathbf{v}_{obs})(\mathbf{\hat{n}}_{j}\cdot\mathbf{v}_{obs})\nonumber\\
&&-\epsilon^{2}\sum_{j,h=1}^{2}v^{2}_{obs}\mathbf{\hat{n}}_{j}\cdot\frac{1}{c}\delta\dot{\mathbf{x}}_{1PN}(\mathbf{x}_{N})|_{h}
+3\epsilon^{2}\sum_{j=1}^{2}(\mathbf{\hat{n}}_{1}\cdot\mathbf{v}_{obs})
(\mathbf{\hat{n}}_{2}\cdot\mathbf{v}_{obs})\mathbf{\hat{n}}_{j}\cdot\frac{1}{c}\delta\dot{\mathbf{x}}_{1PN}(\mathbf{x}_{N})|_{j}
\nonumber\\
&&
+3\epsilon^{2}\sum_{j=1}^{2}(\mathbf{\hat{n}}_{j}\cdot\mathbf{v}_{obs})^{2}\mathbf{\hat{n}}_{j}\cdot\frac{1}{c}\delta\dot{\mathbf{x}}_{1PN}(\mathbf{x}_{N})|_{j}
+\epsilon^{2}\sum_{j,h,l=1}^{2}\sum_{j\neq
h}(\mathbf{\hat{n}}_{l}\cdot\mathbf{v}_{obs})^{2}\mathbf{\hat{n}}_{j}\cdot\frac{1}{c}\delta\dot{\mathbf{x}}_{1PN}(\mathbf{x}_{N})|_{h}
\nonumber\\
&&+\epsilon^{2}\sum_{j,h=1}^{2}\sum_{j\neq
h}(\mathbf{\hat{n}}_{j}\cdot\mathbf{v}_{obs})^{2}\mathbf{\hat{n}}_{h}\cdot\frac{1}{c}\delta\dot{\mathbf{x}}_{1PN}(\mathbf{x}_{N})|_{h}\nonumber\\
&&+\epsilon^{2}\sum_{j,h=1}^{2}\sum_{j\neq
h}(\mathbf{\hat{n}}_{1}\cdot\mathbf{v}_{obs})(\mathbf{\hat{n}}_{2}\cdot\mathbf{v}_{obs})\mathbf{\hat{n}}_{j}\cdot\frac{1}{c}\delta\dot{\mathbf{x}}_{1PN}(\mathbf{x}_{N})|_{h}
\nonumber\\
&&-2\epsilon^{2}\sum_{j=1}^{2}(\mathbf{\hat{n}}_{j}\cdot\mathbf{v}_{obs})\mathbf{v}_{obs}\cdot\frac{1}{c}\delta\dot{\mathbf{x}}_{1PN}(\mathbf{x}_{N})|_{j}
-\epsilon^{2}\sum_{j,h=1}^{2}\sum_{j\neq
h}(\mathbf{\hat{n}}_{j}\cdot\mathbf{v}_{obs})\mathbf{v}_{obs}\cdot\frac{1}{c}\delta\dot{\mathbf{x}}_{1PN}(\mathbf{x}_{N})|_{h}
\nonumber\\
&&+2\epsilon^{2}\sum_{j=1}^{2}v^{2}_{obs}(\mathbf{\hat{n}}_{1}\cdot\mathbf{\hat{n}}_{2})\mathbf{\hat{n}}_{j}\cdot\frac{1}{c}\delta\dot{\mathbf{x}}_{1PN}(\mathbf{x}_{N})|_{j}
+2\epsilon^{2}\sum_{j=1}^{2}(\mathbf{\hat{n}}_{1}\cdot\mathbf{\hat{n}}_{2})(\mathbf{\hat{n}}_{j}\cdot\mathbf{v}_{obs})\mathbf{v}_{obs}\cdot\frac{1}{c}\delta\dot{\mathbf{x}}_{1PN}(\mathbf{x}_{N})|_{j}
\nonumber\\
&&+\epsilon^{2}\sum_{j,h=1}^{2}\sum_{j\neq
h}(\mathbf{\hat{n}}_{1}\cdot\mathbf{\hat{n}}_{2})(\mathbf{\hat{n}}_{j}\cdot\mathbf{v}_{obs})
\mathbf{v}_{obs}\cdot\frac{1}{c}\delta\dot{\mathbf{x}}_{1PN}(\mathbf{x}_{N})|_{h}
\nonumber\\
&&-\epsilon^{2}\sum_{j,h=1}^{2}\sum_{j\neq
h}(\mathbf{\hat{n}}_{1}\cdot\mathbf{\hat{n}}_{2})(\mathbf{\hat{n}}_{j}\cdot\mathbf{v}_{obs})^{2}
\mathbf{\hat{n}}_{h}\cdot\frac{1}{c}\delta\dot{\mathbf{x}}_{1PN}(\mathbf{x}_{N})|_{h}
\nonumber\\
&&-3\epsilon^{2}\sum_{j=1}^{2}(\mathbf{\hat{n}}_{1}\cdot\mathbf{v}_{obs})
(\mathbf{\hat{n}}_{2}\cdot\mathbf{v}_{obs})(\mathbf{\hat{n}}_{1}\cdot\mathbf{\hat{n}}_{2})\mathbf{\hat{n}}_{j}\cdot\frac{1}{c}\delta\dot{\mathbf{x}}_{1PN}(\mathbf{x}_{N})|_{j}
\nonumber\\
&&-6\epsilon^{2}\sum_{j=1}^{2}(\mathbf{\hat{n}}_{1}\cdot\mathbf{\hat{n}}_{2})(\mathbf{\hat{n}}_{j}\cdot\mathbf{v}_{obs})^{2}\mathbf{\hat{n}}_{j}\cdot\frac{1}{c}\delta\dot{\mathbf{x}}_{1PN}(\mathbf{x}_{N})|_{j},
\end{eqnarray}
\begin{eqnarray}
\overset{(4)}{f}_{2PN}&=&\frac{1}{2}\eta\epsilon^{4}\sum_{k,j=1}^{2}\sum_{k\neq
j}\frac{G^{2}m^{2}_{\odot}}{r^{4}_{obs\odot}}(\mathbf{\hat{n}}_{k}\cdot\mathbf{r}_{obs\odot})(\mathbf{\hat{n}}_{j}\cdot\mathbf{r}_{obs\odot})
-\frac{1}{2}\eta\epsilon^{4}\sum_{k=1}^{2}\frac{G^{2}m^{2}_{\odot}}{r^{4}_{obs\odot}}(\mathbf{\hat{n}}_{k}\cdot\mathbf{r}_{obs\odot})^{2}(\mathbf{\hat{n}}_{1}\cdot\mathbf{\hat{n}}_{2})\nonumber\\
&&-\sum_{j,l,n=1}^{2}\sum_{l\neq
n}\mathbf{\hat{n}}_{j}\cdot\frac{1}{c}\delta\dot{\mathbf{x}}_{1PN}(\mathbf{x}_{N})|_{j}\mathbf{\hat{n}}_{l}\cdot\frac{1}{c}\delta\dot{\mathbf{x}}_{1PN}(\mathbf{x}_{N})|_{n}\nonumber\\
&&-\frac{1}{2}\sum_{j=1}^{2}(\mathbf{\hat{n}}_{1}\cdot\mathbf{\hat{n}}_{2})\frac{1}{c}\delta\dot{\mathbf{x}}_{1PN}(\mathbf{x}_{N})|_{j}\cdot\frac{1}{c}\delta\dot{\mathbf{x}}_{1PN}(\mathbf{x}_{N})|_{j}
\nonumber\\
&&+\frac{3}{2}\sum_{j=1}^{2}(\mathbf{\hat{n}}_{1}\cdot\mathbf{\hat{n}}_{2})
\mathbf{\hat{n}}_{j}\cdot\frac{1}{c}\delta\dot{\mathbf{x}}_{1PN}(\mathbf{x}_{N})|_{j}\mathbf{\hat{n}}_{j}\cdot\frac{1}{c}\delta\dot{\mathbf{x}}_{1PN}(\mathbf{x}_{N})|_{j}
+\frac{1}{c}\delta\dot{\mathbf{x}}_{1PN}(\mathbf{x}_{N})|_{1}\cdot\frac{1}{c}\delta\dot{\mathbf{x}}_{1PN}(\mathbf{x}_{N})|_{2}
\nonumber\\
&&+(\mathbf{\hat{n}}_{1}\cdot\mathbf{\hat{n}}_{2})\mathbf{\hat{n}}_{2}\cdot\frac{1}{c}\delta\dot{\mathbf{x}}_{1PN}(\mathbf{x}_{N})|_{2}\mathbf{\hat{n}}_{1}\cdot\frac{1}{c}\delta\dot{\mathbf{x}}_{1PN}(\mathbf{x}_{N})|_{1}
+\sum_{j,h=1}^{2}\sum_{j\neq
h}\mathbf{\hat{n}}_{j}\cdot\frac{1}{c}\delta\dot{\mathbf{x}}_{2PN}(\mathbf{x}_{N})|_{h}\nonumber\\
&&-\sum_{j=1}^{2}(\mathbf{\hat{n}}_{1}\cdot\mathbf{\hat{n}}_{2})\mathbf{\hat{n}}_{j}\cdot\frac{1}{c}\delta\dot{\mathbf{x}}_{2PN}(\mathbf{x}_{N})|_{j},
\end{eqnarray}
where $\bm{\mathcal{C}}(\mathbf{x})|_{1}$ and
$\bm{\mathcal{C}}(\mathbf{x})|_{2}$ denote that this function
$\bm{\mathcal{C}}(\mathbf{x})$ takes respectively the value at the
signal 1 and the signal 2.

\section{Angle expressions \label{appendixD}}

We list the expressions of the deflection angles in Eq.
(\ref{theta}).
\begin{eqnarray}
\overset{(1)}{\vartheta}_{obs}&=&\epsilon[(\mathbf{\hat{n}}_{1}\cdot\mathbf{v}_{obs})
+(\mathbf{\hat{n}}_{2}\cdot\mathbf{v}_{obs})]\frac{1-\cos\vartheta_{0}}{\sin\vartheta_{0}},
\end{eqnarray}
\begin{eqnarray}
\overset{(2)}{\vartheta}_{obs}&=&\epsilon^{2}\frac{\sin\vartheta_{0}}{(1+\cos\vartheta_{0})^{2}}\bigg[
\bigg((\mathbf{\hat{n}}_{1}\cdot\mathbf{v}_{obs})^{2}+(\mathbf{\hat{n}}_{2}\cdot\mathbf{v}_{obs})^{2}\bigg)\bigg(1+\frac{1}{2}\cos\vartheta_{0}\bigg)\nonumber\\
&&+(\mathbf{\hat{n}}_{1}\cdot\mathbf{v}_{obs})(\mathbf{\hat{n}}_{2}\cdot\mathbf{v}_{obs})
-v^{2}_{obs}(1+\cos\vartheta_{0})\bigg],
\end{eqnarray}
\begin{eqnarray}
\overset{(2)}{\vartheta}_{1PN}&=&(1+\gamma)\epsilon^{2}\sum_{A}\frac{Gm_{A}}{r_{A}\sin\vartheta_{0}}\bigg\{
\frac{\mathbf{\hat{n}}_{1}\cdot[\mathbf{\hat{n}}_{2}\times(\mathbf{r}_{A}\times\mathbf{\hat{n}}_{2})]}{r_{A}-\mathbf{\hat{n}}_{2}\cdot\mathbf{r}_{A}}
+\frac{\mathbf{\hat{n}}_{2}\cdot[\mathbf{\hat{n}}_{1}\times(\mathbf{r}_{A}\times\mathbf{\hat{n}}_{1})]}{r_{A}-\mathbf{\hat{n}}_{1}\cdot\mathbf{r}_{A}}\bigg\},
\end{eqnarray}
\begin{eqnarray}
\overset{(2)}{\vartheta}_{Q}&=&-(1+\gamma)\epsilon^{2}\sum_{A}\frac{G}{r^{3}_{A}\sin\vartheta_{0}}\bigg\{
J^{<ik>}_{A}[\hat{n}^{i}_{1}-(\mathbf{\hat{n}}_{1}\cdot\mathbf{\hat{n}}_{2})\hat{n}^{i}_{2}]\bigg[\frac{2r_{A}-\mathbf{\hat{n}}_{2}\cdot\mathbf{r}_{A}}{(r_{A}-\mathbf{\hat{n}}_{2}\cdot\mathbf{r}_{A})^{2}}d^{k}_{2A}-\hat{n}^{k}_{2}\bigg]\nonumber\\
&&+\frac{3J^{<jk>}_{A}}{2r^{2}_{A}}(\mathbf{\hat{n}}_{1}\cdot\mathbf{d}_{2A})\bigg[2\hat{n}_{2}^{j}d^{k}_{2A}
+(\mathbf{\hat{n}}_{2}\cdot\mathbf{r}_{A})\hat{n}^{j}_{2}\hat{n}^{k}_{2}-(\mathbf{\hat{n}}_{2}\cdot\mathbf{r}_{A})\frac{d^{j}_{2A}d^{k}_{2A}}{d^{2}_{2A}}\bigg]\nonumber\\
&&+\frac{3J^{<jk>}_{A}}{2r^{2}_{A}}(\mathbf{\hat{n}}_{2}\cdot\mathbf{d}_{1A})\bigg[2\hat{n}_{1}^{j}d^{k}_{1A}
+(\mathbf{\hat{n}}_{1}\cdot\mathbf{r}_{A})\hat{n}^{j}_{1}\hat{n}^{k}_{1}-(\mathbf{\hat{n}}_{1}\cdot\mathbf{r}_{A})\frac{d^{j}_{1A}d^{k}_{1A}}{d^{2}_{1A}}\bigg]\nonumber\\
&&+J^{<ik>}_{A}[\hat{n}^{i}_{2}-(\mathbf{\hat{n}}_{1}\cdot\mathbf{\hat{n}}_{2})\hat{n}^{i}_{1}]\bigg[\frac{2r_{A}-\mathbf{\hat{n}}_{1}\cdot\mathbf{r}_{A}}{(r_{A}-\mathbf{\hat{n}}_{1}\cdot\mathbf{r}_{A})^{2}}d^{k}_{1A}-\hat{n}^{k}_{1}\bigg]\bigg\},
\end{eqnarray}
\begin{eqnarray}
\overset{(3)}{\vartheta}_{obs}&=&\epsilon^{3}\frac{(1+2\cos^{2}\vartheta_{0})(1-\cos\vartheta_{0})^{3}}{6\sin^{5}\vartheta_{0}}
\bigg[(\mathbf{\hat{n}}_{1}\cdot\mathbf{v}_{obs})+(\mathbf{\hat{n}}_{2}\cdot\mathbf{v}_{obs})\bigg]^{3}\nonumber\\
&&+\epsilon^{3}\frac{\cos\vartheta_{0}(1-\cos\vartheta_{0})^{2}}{\sin^{3}\vartheta_{0}}
[(\mathbf{\hat{n}}_{1}\cdot\mathbf{v}_{obs})+(\mathbf{\hat{n}}_{2}\cdot\mathbf{v}_{obs})]\bigg[v^{2}_{obs}
-(\mathbf{\hat{n}}_{1}\cdot\mathbf{v}_{obs})^{2}-(\mathbf{\hat{n}}_{2}\cdot\mathbf{v}_{obs})^{2}\nonumber\\
&&
-(\mathbf{\hat{n}}_{1}\cdot\mathbf{v}_{obs})(\mathbf{\hat{n}}_{2}\cdot\mathbf{v}_{obs})\bigg]
+\epsilon^{3}\frac{1-\cos\vartheta_{0}}{\sin\vartheta_{0}}
\bigg[(\mathbf{\hat{n}}_{1}\cdot\mathbf{v}_{obs})^{3}+(\mathbf{\hat{n}}_{2}\cdot\mathbf{v}_{obs})^{3}\nonumber\\
&&-(\mathbf{\hat{n}}_{1}\cdot\mathbf{v}_{obs})v^{2}_{obs}-(\mathbf{\hat{n}}_{2}\cdot\mathbf{v}_{obs})v^{2}_{obs}
+(\mathbf{\hat{n}}_{1}\cdot\mathbf{v}_{obs})^{2}(\mathbf{\hat{n}}_{2}\cdot\mathbf{v}_{obs})+(\mathbf{\hat{n}}_{2}\cdot\mathbf{v}_{obs})^{2}(\mathbf{\hat{n}}_{1}\cdot\mathbf{v}_{obs})\bigg]\nonumber\\
&&+(1+\gamma)\epsilon^{3}\frac{1-\cos\vartheta_{0}}{\sin\vartheta_{0}}
\sum_{A}\frac{Gm_{A}}{r_{A}}\bigg[(\mathbf{\hat{n}}_{1}\cdot\mathbf{v}_{obs})+(\mathbf{\hat{n}}_{2}\cdot\mathbf{v}_{obs})
-\frac{\mathbf{v}_{obs}\cdot\mathbf{d}_{1A}}{r_{A}-\mathbf{\hat{n}}_{1}\cdot\mathbf{r}_{A}}
-\frac{\mathbf{v}_{obs}\cdot\mathbf{d}_{2A}}{r_{A}-\mathbf{\hat{n}}_{2}\cdot\mathbf{r}_{A}}\bigg]\nonumber\\
&&+(1+\gamma)\epsilon^{3}\frac{1}{\sin\vartheta_{0}}\sum_{A}\frac{Gm_{A}}{r_{A}}[(\mathbf{\hat{n}}_{1}\cdot\mathbf{v}_{obs})+(\mathbf{\hat{n}}_{2}\cdot\mathbf{v}_{obs})]
\bigg[\frac{\mathbf{\hat{n}}_{1}\cdot\mathbf{d}_{2A}}{r_{A}-\mathbf{\hat{n}_{2}\cdot\mathbf{r}_{A}}}
+\frac{\mathbf{\hat{n}}_{2}\cdot\mathbf{d}_{1A}}{r_{A}-\mathbf{\hat{n}_{1}\cdot\mathbf{r}_{A}}}\bigg],
\end{eqnarray}
\begin{eqnarray}
\overset{(3)}{\vartheta}_{OM}&=&(1+\gamma)\epsilon^{3}\sum_{A}\frac{Gm_{A}}{\sin\vartheta_{0}}\bigg[
-\frac{(\mathbf{\hat{n}}_{1}\cdot\mathbf{d}_{2A})(\mathbf{\hat{n}}_{2}\cdot\mathbf{v}_{A})}{r_{A}(r_{A}-\mathbf{\hat{n}}_{2}\cdot\mathbf{r}_{A})}
-\frac{(\mathbf{\hat{n}}_{2}\cdot\mathbf{d}_{1A})(\mathbf{\hat{n}}_{1}\cdot\mathbf{v}_{A})}{r_{A}(r_{A}-\mathbf{\hat{n}}_{1}\cdot\mathbf{r}_{A})}
+\frac{(\mathbf{\hat{n}}_{1}\cdot\mathbf{d}_{2A})(\mathbf{\hat{n}}_{2}\cdot\mathbf{v}_{A})}{(r_{A}-\mathbf{\hat{n}}_{2}\cdot\mathbf{r}_{A})^{2}}\nonumber\\
&&+\frac{(\mathbf{\hat{n}}_{2}\cdot\mathbf{d}_{1A})(\mathbf{\hat{n}}_{1}\cdot\mathbf{v}_{A})}{(r_{A}-\mathbf{\hat{n}}_{1}\cdot\mathbf{r}_{A})^{2}}
-\frac{(\mathbf{\hat{n}}_{1}\cdot\mathbf{d}_{2A})(\mathbf{r}_{A}\cdot\mathbf{v}_{A})}{r_{A}(r_{A}-\mathbf{\hat{n}}_{2}\cdot\mathbf{r}_{A})^{2}}
-\frac{(\mathbf{\hat{n}}_{2}\cdot\mathbf{d}_{1A})(\mathbf{r}_{A}\cdot\mathbf{v}_{A})}{r_{A}(r_{A}-\mathbf{\hat{n}}_{1}\cdot\mathbf{r}_{A})^{2}}\nonumber\\
&&-\frac{(\mathbf{\hat{n}}_{1}\cdot\mathbf{r}_{A})(\mathbf{\hat{n}}_{2}\cdot\mathbf{v}_{A})}{r_{A}(r_{A}-\mathbf{\hat{n}}_{2}\cdot\mathbf{r}_{A})}
-\frac{(\mathbf{\hat{n}}_{2}\cdot\mathbf{r}_{A})(\mathbf{\hat{n}}_{1}\cdot\mathbf{v}_{A})}{r_{A}(r_{A}-\mathbf{\hat{n}}_{1}\cdot\mathbf{r}_{A})}
+\frac{(\mathbf{\hat{n}}_{2}\cdot\mathbf{r}_{A})(\mathbf{\hat{n}}_{1}\cdot\mathbf{v}_{A})}{r_{A}(r_{A}-\mathbf{\hat{n}}_{2}\cdot\mathbf{r}_{A})}
+\frac{(\mathbf{\hat{n}}_{1}\cdot\mathbf{r}_{A})(\mathbf{\hat{n}}_{2}\cdot\mathbf{v}_{A})}{r_{A}(r_{A}-\mathbf{\hat{n}}_{1}\cdot\mathbf{r}_{A})}\nonumber\\
&&-\frac{\mathbf{\hat{n}}_{1}\cdot\mathbf{v}_{A}}{r_{A}}
-\frac{\mathbf{\hat{n}}_{2}\cdot\mathbf{v}_{A}}{r_{A}}
+(\mathbf{\hat{n}}_{1}\cdot\mathbf{\hat{n}}_{2})\frac{\mathbf{\hat{n}}_{1}\cdot\mathbf{v}_{A}}{r_{A}}
+(\mathbf{\hat{n}}_{1}\cdot\mathbf{\hat{n}}_{2})\frac{\mathbf{\hat{n}}_{2}\cdot\mathbf{v}_{A}}{r_{A}}\bigg],
\end{eqnarray}
\begin{eqnarray}
\overset{(3)}{\vartheta}_{S}&=&-(1+\gamma)\epsilon^{3}\sum_{A}\frac{G}{\sin\vartheta_{0}}\bigg\{-
\frac{\mathbf{S}_{A}\cdot(\mathbf{\hat{n}}_{1}\times\mathbf{\hat{n}}_{2})]}{d^{2}_{2A}}\bigg[1+\frac{(\mathbf{\hat{n}}_{2}\cdot\mathbf{r}_{A})}{r_{A}}\bigg]
\bigg[\frac{(\mathbf{\hat{n}}_{2}\cdot\mathbf{r}_{A})}{r_{A}}
-\frac{(\mathbf{\hat{n}}_{2}\cdot\mathbf{r}_{A})^{2}}{r^{2}_{A}}+1\bigg]\nonumber\\
&&+\frac{\mathbf{\hat{n}}_{1}\cdot(\mathbf{S}_{A}\times\mathbf{d}_{2A})}{r^{3}_{A}}
-
\frac{\mathbf{S}_{A}\cdot(\mathbf{\hat{n}}_{2}\times\mathbf{\hat{n}}_{1})]}{d^{2}_{1A}}\bigg[1+\frac{(\mathbf{\hat{n}}_{1}\cdot\mathbf{r}_{A})}{r_{A}}\bigg]
\bigg[\frac{(\mathbf{\hat{n}}_{1}\cdot\mathbf{r}_{A})}{r_{A}}
-\frac{(\mathbf{\hat{n}}_{1}\cdot\mathbf{r}_{A})^{2}}{r^{2}_{A}}+1\bigg]\nonumber\\
&&-\frac{\mathbf{\hat{n}}_{2}\cdot\mathbf{d}_{1A}}{d^{4}_{1A}}[\mathbf{S}_{A}\cdot(\mathbf{\hat{n}}_{1}\times\mathbf{d}_{1A})]
\bigg[2-\frac{\mathbf{\hat{n}}_{1}\cdot\mathbf{r}_{A}}{r_{A}}\bigg]
\bigg[1+2\frac{\mathbf{\hat{n}}_{1}\cdot\mathbf{r}_{A}}{r_{A}}+\frac{(\mathbf{\hat{n}}_{1}\cdot\mathbf{r}_{A})^{2}}{r^{2}_{A}}\bigg]\nonumber\\
&&-\frac{\mathbf{\hat{n}}_{1}\cdot\mathbf{d}_{2A}}{d^{4}_{2A}}[\mathbf{S}_{A}\cdot(\mathbf{\hat{n}}_{2}\times\mathbf{d}_{2A})]
\bigg[2-\frac{\mathbf{\hat{n}}_{2}\cdot\mathbf{r}_{A}}{r_{A}}\bigg]
\bigg[1+2\frac{\mathbf{\hat{n}}_{2}\cdot\mathbf{r}_{A}}{r_{A}}+\frac{(\mathbf{\hat{n}}_{2}\cdot\mathbf{r}_{A})^{2}}{r^{2}_{A}}\bigg]\nonumber\\
&&
+\frac{\mathbf{\hat{n}}_{2}\cdot(\mathbf{S}_{A}\times\mathbf{d}_{1A})}{r^{3}_{A}}
-(\mathbf{\hat{n}}_{1}\cdot\mathbf{\hat{n}}_{2})\frac{\mathbf{\hat{n}}_{1}\cdot(\mathbf{S}_{A}\times\mathbf{d}_{1A})}{r^{3}_{A}}
-(\mathbf{\hat{n}}_{1}\cdot\mathbf{\hat{n}}_{2})\frac{\mathbf{\hat{n}}_{2}\cdot(\mathbf{S}_{A}\times\mathbf{d}_{2A})}{r^{3}_{A}}
\bigg\},
\end{eqnarray}
\begin{eqnarray}
\overset{(4)}{\vartheta}_{obs}&=&-\epsilon^{4}\frac{\cos\vartheta_{0}(3+2\cos^{2}\vartheta_{0})(1-\cos\vartheta_{0})^{4}}{8\sin^{7}\vartheta_{0}}
[(\mathbf{\hat{n}}_{1}\cdot\mathbf{v}_{obs})+(\mathbf{\hat{n}}_{2}\cdot\mathbf{v}_{obs})]^{4}\nonumber\\
&&+\epsilon^{4}\frac{(1+2\cos^{2}\vartheta_{0})(1-\cos\vartheta_{0})^{3}}{2\sin^{5}\vartheta_{0}}[(\mathbf{\hat{n}}_{1}\cdot\mathbf{v}_{obs})+(\mathbf{\hat{n}}_{2}\cdot\mathbf{v}_{obs})]^{2}
[(\mathbf{\hat{n}}_{1}\cdot\mathbf{v}_{obs})^{2}+(\mathbf{\hat{n}}_{2}\cdot\mathbf{v}_{obs})^{2}\nonumber\\
&&+(\mathbf{\hat{n}}_{1}\cdot\mathbf{v}_{obs})(\mathbf{\hat{n}}_{2}\cdot\mathbf{v}_{obs})-v^{2}_{obs}]
-\epsilon^{4}\frac{\cos\vartheta_{0}(1-\cos\vartheta_{0})^{2}}{\sin^{3}\vartheta_{0}}[(\mathbf{\hat{n}}_{1}\cdot\mathbf{v}_{obs})+(\mathbf{\hat{n}}_{2}\cdot\mathbf{v}_{obs})]\nonumber\\
&&\times[(\mathbf{\hat{n}}_{1}\cdot\mathbf{v}_{obs})^{3}+(\mathbf{\hat{n}}_{2}\cdot\mathbf{v}_{obs})^{3}
-(\mathbf{\hat{n}}_{1}\cdot\mathbf{v}_{obs})v^{2}_{o}-(\mathbf{\hat{n}}_{2}\cdot\mathbf{v}_{obs})v^{2}_{obs}
+(\mathbf{\hat{n}}_{1}\cdot\mathbf{v}_{obs})^{2}(\mathbf{\hat{n}}_{2}\cdot\mathbf{v}_{obs})\nonumber\\
&&+(\mathbf{\hat{n}}_{2}\cdot\mathbf{v}_{obs})^{2}(\mathbf{\hat{n}}_{1}\cdot\mathbf{v}_{obs})]
-\epsilon^{4}\frac{\cos\vartheta_{0}(1-\cos\vartheta_{0})^{2}}{2\sin^{3}\vartheta_{0}}
[(\mathbf{\hat{n}}_{1}\cdot\mathbf{v}_{obs})^{2}+(\mathbf{\hat{n}}_{2}\cdot\mathbf{v}_{obs})^{2}\nonumber\\
&&+(\mathbf{\hat{n}}_{1}\cdot\mathbf{v}_{obs})(\mathbf{\hat{n}}_{2}\cdot\mathbf{v}_{obs})
-v^{2}_{obs}]^{2}
+\epsilon^{4}\frac{1-\cos\vartheta_{0}}{\sin\vartheta_{0}}[(\mathbf{\hat{n}}_{1}\cdot\mathbf{v}_{obs})^{4}+(\mathbf{\hat{n}}_{2}\cdot\mathbf{v}_{obs})^{4}\nonumber\\
&&-(\mathbf{\hat{n}}_{1}\cdot\mathbf{v}_{obs})^{2}v^{2}_{obs}-(\mathbf{\hat{n}}_{2}\cdot\mathbf{v}_{obs})^{2}v^{2}_{obs}
-(\mathbf{\hat{n}}_{1}\cdot\mathbf{v}_{obs})(\mathbf{\hat{n}}_{2}\cdot\mathbf{v}_{obs})v^{2}_{obs}
+(\mathbf{\hat{n}}_{1}\cdot\mathbf{v}_{obs})^{3}(\mathbf{\hat{n}}_{2}\cdot\mathbf{v}_{obs})\nonumber\\
&&+(\mathbf{\hat{n}}_{2}\cdot\mathbf{v}_{obs})^{3}(\mathbf{\hat{n}}_{1}\cdot\mathbf{v}_{obs})
+(\mathbf{\hat{n}}_{1}\cdot\mathbf{v}_{obs})^{2}(\mathbf{\hat{n}}_{2}\cdot\mathbf{v}_{obs})^{2}]\nonumber\\
&&-(1+\gamma)\epsilon^{4}\frac{\cos\vartheta_{0}(1-\cos\vartheta_{0})^{2}}{\sin^{3}\vartheta_{0}}
\frac{Gm_{\odot}}{r_{\odot}}[(\mathbf{\hat{n}}_{1}\cdot\mathbf{v}_{obs})+(\mathbf{\hat{n}}_{2}\cdot\mathbf{v}_{obs})]\bigg[(\mathbf{\hat{n}}_{1}\cdot\mathbf{v}_{obs})+(\mathbf{\hat{n}}_{2}\cdot\mathbf{v}_{obs})\nonumber\\
&&-\frac{\mathbf{v}_{obs}\cdot\mathbf{d}_{1\odot}}{r_{\odot}-\mathbf{\hat{n}}_{1}\cdot\mathbf{r}_{\odot}}
-\frac{\mathbf{v}_{obs}\cdot\mathbf{d}_{2\odot}}{r_{\odot}-\mathbf{\hat{n}}_{2}\cdot\mathbf{r}_{\odot}}\bigg]
+(1+\gamma)\epsilon^{4}\frac{(1+2\cos^{2}\vartheta_{0}-3\cos\vartheta_{0})(1-\cos\vartheta_{0})}{2\sin^{5}\vartheta_{0}}\frac{Gm_{\odot}}{r_{\odot}}
\nonumber\\
&&\times[(\mathbf{\hat{n}}_{1}\cdot\mathbf{v}_{obs})+(\mathbf{\hat{n}}_{2}\cdot\mathbf{v}_{obs})]^{2}\bigg[\frac{\mathbf{\hat{n}}_{1}\cdot\mathbf{d}_{2\odot}}{r_{\odot}-\mathbf{\hat{n}_{2}\cdot\mathbf{r}_{\odot}}}
+\frac{\mathbf{\hat{n}}_{2}\cdot\mathbf{d}_{1\odot}}{r_{\odot}-\mathbf{\hat{n}_{1}\cdot\mathbf{r}_{\odot}}}\bigg]
+(1+\gamma)\epsilon^{4}\frac{(\cos\vartheta_{0}-1)\cos\vartheta_{0}}{\sin^{3}\vartheta_{0}}\frac{Gm_{\odot}}{r_{\odot}}\nonumber\\
&&\times[(\mathbf{\hat{n}}_{1}\cdot\mathbf{v}_{obs})^{2}+(\mathbf{\hat{n}}_{2}\cdot\mathbf{v}_{obs})^{2}
+(\mathbf{\hat{n}}_{1}\cdot\mathbf{v}_{obs})(\mathbf{\hat{n}}_{2}\cdot\mathbf{v}_{obs})-v^{2}_{obs}]
\bigg[\frac{\mathbf{\hat{n}}_{1}\cdot\mathbf{d}_{2\odot}}{r_{\odot}-\mathbf{\hat{n}_{2}\cdot\mathbf{r}_{\odot}}}
+\frac{\mathbf{\hat{n}}_{2}\cdot\mathbf{d}_{1\odot}}{r_{\odot}-\mathbf{\hat{n}_{1}\cdot\mathbf{r}_{\odot}}}\bigg]\nonumber\\
&&+(1+\gamma)\epsilon^{4}\cot\vartheta_{0}\frac{Gm_{\odot}}{r_{\odot}}\bigg[2v^{2}_{obs}-2(\mathbf{\hat{n}}_{1}\cdot\mathbf{v}_{obs})^{2}
-2(\mathbf{\hat{n}}_{2}\cdot\mathbf{v}_{obs})^{2}-2(\mathbf{\hat{n}}_{1}\cdot\mathbf{v}_{obs})(\mathbf{\hat{n}}_{2}\cdot\mathbf{v}_{obs})\nonumber\\
&&+\frac{(\mathbf{\hat{n}}_{2}\cdot\mathbf{v}_{obs})(\mathbf{v}_{obs}\cdot\mathbf{d}_{1\odot})}{r_{\odot}-\mathbf{\hat{n}}_{1}\cdot\mathbf{r}_{\odot}}
+\frac{(\mathbf{\hat{n}}_{1}\cdot\mathbf{v}_{obs})(\mathbf{v}_{obs}\cdot\mathbf{d}_{2\odot})}{r_{\odot}-\mathbf{\hat{n}}_{2}\cdot\mathbf{r}_{\odot}}
+2\frac{(\mathbf{\hat{n}}_{2}\cdot\mathbf{v}_{obs})(\mathbf{v}_{obs}\cdot\mathbf{d}_{2\odot})}{r_{\odot}-\mathbf{\hat{n}}_{2}\cdot\mathbf{r}_{\odot}}\nonumber\\
&&+2\frac{(\mathbf{\hat{n}}_{1}\cdot\mathbf{v}_{obs})(\mathbf{v}_{obs}\cdot\mathbf{d}_{1\odot})}{r_{\odot}-\mathbf{\hat{n}}_{1}\cdot\mathbf{r}_{\odot}}\bigg]
+(1+\gamma)\epsilon^{4}\frac{1}{\sin\vartheta_{0}}\frac{Gm_{\odot}}{r_{\odot}}[(\mathbf{\hat{n}}_{1}\cdot\mathbf{v}_{obs})(\mathbf{\hat{n}}_{2}\cdot\mathbf{v}_{obs})
-v^{2}_{obs}+(\mathbf{\hat{n}}_{1}\cdot\mathbf{v}_{obs})^{2}\nonumber\\
&&+(\mathbf{\hat{n}}_{2}\cdot\mathbf{v}_{obs})^{2}]\bigg(2+
\frac{\mathbf{\hat{n}}_{1}\cdot\mathbf{d}_{2\odot}}{r_{\odot}-\mathbf{\hat{n}}_{2}\cdot\mathbf{r}_{\odot}}
+\frac{\mathbf{\hat{n}}_{2}\cdot\mathbf{d}_{1\odot}}{r_{\odot}-\mathbf{\hat{n}}_{1}\cdot\mathbf{r}_{\odot}}\bigg)\nonumber\\
&&-(1+\gamma)\epsilon^{4}\frac{1}{\sin\vartheta_{0}}\frac{Gm_{\odot}}{r_{\odot}}\bigg[
2\frac{(\mathbf{\hat{n}}_{1}\cdot\mathbf{v}_{obs})(\mathbf{v}_{obs}\cdot\mathbf{d}_{1\odot})}{r_{\odot}-\mathbf{\hat{n}}_{1}\cdot\mathbf{r}_{\odot}}
+2\frac{(\mathbf{\hat{n}}_{2}\cdot\mathbf{v}_{obs})(\mathbf{v}_{obs}\cdot\mathbf{d}_{2\odot})}{r_{\odot}-\mathbf{\hat{n}}_{2}\cdot\mathbf{r}_{\odot}}\nonumber\\
&&+\frac{(\mathbf{\hat{n}}_{1}\cdot\mathbf{v}_{obs})(\mathbf{v}_{obs}\cdot\mathbf{d}_{2\odot})}{r_{\odot}-\mathbf{\hat{n}}_{2}\cdot\mathbf{r}_{\odot}}
+\frac{(\mathbf{\hat{n}}_{2}\cdot\mathbf{v}_{obs})(\mathbf{v}_{obs}\cdot\mathbf{d}_{1\odot})}{r_{\odot}-\mathbf{\hat{n}}_{1}\cdot\mathbf{r}_{\odot}}\bigg],
\end{eqnarray}
\begin{eqnarray}
\overset{(4)}{\vartheta}_{2PN}&=&-(1+\gamma)^{2}\epsilon^{4}\frac{G^{2}m^{2}_{\odot}}{\sin\vartheta_{0}}\bigg[
\frac{\mathbf{\hat{n}}_{2}\cdot\mathbf{d}_{1\odot}}{d^{3}_{1\odot}}\bigg(1+\frac{\mathbf{\hat{n}}_{1}\cdot\mathbf{r}_{\odot}}{r_{\odot}}\bigg)
+\frac{\mathbf{\hat{n}}_{1}\cdot\mathbf{d}_{2\odot}}{d^{3}_{2\odot}}\bigg(1+\frac{\mathbf{\hat{n}}_{2}\cdot\mathbf{r}_{\odot}}{r_{\odot}}\bigg)\bigg]\nonumber\\
&&-[\beta-\frac{1}{4}(2\varsigma+\eta)+\gamma^{2}-1]\epsilon^{4}\frac{G^{2}m^{2}_{\odot}}{r^{2}_{\odot}}
\bigg[\frac{(\mathbf{\hat{n}}_{1}\cdot\mathbf{r}_{\odot})}{d_{1\odot}}
+\frac{(\mathbf{\hat{n}}_{2}\cdot\mathbf{r}_{\odot})}{d_{2\odot}}\bigg]\nonumber\\
&&-[\beta-\frac{1}{4}(2\varsigma+\eta)-2(1+\gamma)]\epsilon^{4}\frac{G^{2}m^{2}_{\odot}}{\sin\vartheta_{0}}
\bigg\{\frac{\mathbf{\hat{n}}_{1}\cdot\mathbf{d}_{2\odot}}{d^{3}_{2\odot}}\bigg[\frac{\pi}{2}+\arctan\bigg(\frac{\mathbf{\hat{n}}_{2}\cdot\mathbf{r}_{\odot}}{d_{2\odot}}\bigg)\bigg]\nonumber\\
&&+\frac{\mathbf{\hat{n}}_{2}\cdot\mathbf{d}_{1\odot}}{d^{3}_{1\odot}}\bigg[\frac{\pi}{2}+\arctan\bigg(\frac{\mathbf{\hat{n}}_{1}\cdot\mathbf{r}_{\odot}}{d_{1\odot}}\bigg)\bigg]\bigg\}.
\end{eqnarray}


\end{document}